%% file: 2DFlatBands.tex
\documentclass[longbibliography,aps,prb,reprint,superscriptaddress,longbibliography]{revtex4-2}
\pdfoutput=1
\usepackage{blindtext}
\usepackage{natbib}
\usepackage{wrapfig}  
\usepackage{mathrsfs,amsmath,amssymb,color,bm,mathtools}
\usepackage[makeroom]{cancel}
\usepackage{setspace}
\usepackage{booktabs}
\usepackage[Symbol]{upgreek}
\usepackage{siunitx,booktabs}

\usepackage {hyperref}

\usepackage[english]{babel}

\usepackage{bbold}

\usepackage{floatrow}
\newfloatcommand{capbtabbox}{table}[][\FBwidth]

\usepackage{graphicx} 

\usepackage[english]{babel}

\usepackage{bbold}
\usepackage{siunitx}

\DeclareSymbolFont{ntxletters}{OML}{ntxmi}{m}{it}  
\DeclareMathSymbol{\tau}{\mathord}{ntxletters}{28}  

\newcommand{\subfiglabel}[1]{(#1)}
\newcommand{\figref}[1]{Fig.~\ref{#1}}
\newcommand{\subfigref}[2]{\figref{#1}\subfiglabel{#2}}

\newcommand{\appref}[1]{Appendix~\ref{#1}}
\newcommand{\eqnref}[1]{Eq.~\eqref{#1}}

\newcommand{\tens}{T}

\renewcommand{\bm}{\mathbf}

\newcommand{\papertitle}{Non-singular and singular flat bands in tunable
  acoustic metamaterials}

\makeatletter
\DeclareRobustCommand{\iscircle}{\mathord{\mathpalette\is@circle\relax}}
\newcommand\is@circle[2]{%
	\begingroup
	\sbox\z@{\raisebox{\depth}{$\m@th#1\circ$}}%
	\sbox\tw@{$#1\square$}%
	\resizebox{!}{\ht\tw@}{\usebox{\z@}}%
	\endgroup
}
\makeatother

\begin{document}
	\title{\papertitle}
	\author{Pragalv Karki}
	\email{karki.pragalv@mayo.edu}
	\affiliation{Department of Physics and Institute for Fundamental Science, University of Oregon, Eugene, OR 97403, USA}
	\affiliation{Department of Radiology, Mayo Clinic College of Medicine, Rochester 55905, MN, USA}
	\author{Jayson Paulose}
       	\email{jpaulose@uoregon.edu}
	\affiliation{Department of Physics and Institute for Fundamental Science, University of Oregon, Eugene, OR 97403, USA}
	\affiliation{Material Science Institute, University of Oregon, Eugene, OR 97403, USA}
	
	\begin{abstract}
          Dispersionless flat bands can be classified into two types: (1) non-singular flat bands whose eigenmodes are completely characterized by compact localized states; and (2) singular flat bands that have a discontinuity in their Bloch eigenfunctions at a band touching point with an adjacent dispersive band, thereby requiring additional extended states to span their eigenmode space. In this study, we design and numerically demonstrate two-dimensional thin-plate acoustic metamaterials in which tunable flat bands of both kinds can be achieved.  Non-singular flat bands are achieved by fine-tuning the ratio of the global tension and the bending stiffness in triangular and honeycomb lattices of plate resonators. A singular flat band arises in a kagome lattice due to the underlying lattice geometry, which can be made degenerate with two additional flat bands by tuning the plate tension. A discrete model of the continuum thin-plate system reveals the interplay of geometric and mechanical factors in determining the existence of flat bands of both types. The singular nature of the kagome lattice flat band is established via a metric called the Hilbert-Schmidt distance calculated between a pair of eigenstates infinitesimally close to the quadratic band touching point. We also simulate an acoustic manifestation of a robust boundary mode arising from the singular flat band and protected by real-space topology in a finite system. Our theoretical and computational study establishes a framework for exploring flat-band physics in a tunable classical system, and for designing acoustic metamaterials with potentially useful sound manipulation capabilities. 
     \end{abstract}
	
\maketitle

\section{Introduction}

A flat band is a constant energy or frequency band for all values of the crystal momentum in the Brillouin zone of excitations of a periodic structure. Originally proposed in electronic systems~\cite{Sutherland_1986,Mielke_1991, Tasaki_1994, Tamura_2002, Maksymenko_2012, Hase_2018,Misumi_2017, Zyuzin_2018, Kumar_2021, Morfonios_2021}, flat band models have also been investigated in optical~\cite{Shen_2010, Apaja_2010, Mukherjee_2015, Travkin_2017, Hamidreza_2017, Ge_2018, Longhi_2019} and acoustic\cite{Zheng_2014,Dubois-2019,Wu_2016,Shen-2022} systems as a novel means of manipulating light and sound in artificial structures. Potential applications of flat band physics in optics include lasing~\cite{Longhi_2019}, distortion-free image transmission~\cite{Xia-2016}, logic~\cite{Bastian-2017}, slow-light propagation~\cite{Li-2008}, and mode conversion~\cite{Kim-2022}. Acoustic structures with flat bands enable functionalities such as cloaking~\cite{Zheng_2014}, lensing~\cite{Dubois-2019}, wavefront manipulation~\cite{Wu_2016}, and addressable localized states~\cite{Shen-2022}. These diverse applications primarily exploit the dispersionless character of flat bands and the consequent existence of compact localized states (CLSs)---a set of states belonging to the flat band, each of which is sharply localized with nonzero weight only on a finite subset of sites.

Compact localized states are guaranteed to exist in flat bands arising from lattice models with finite-range interactions between sites~\cite{Rhim_2019,Rhim-Yang_2021}. However, they are not always guaranteed to form a complete spanning set for the space of Bloch eigenfunctions belonging to the flat band. If the flat band touches another band at a particular crystal momentum, the point of band touching can induce a discontinuity in the Bloch eigenfunctions of the flat band when treated as a function of the momentum, which serves as a topological obstruction to finding a spanning set of CLSs~\cite{Bergman2008}. Flat bands can be classified as singular or non-singular based on the presence or absence of such a discontinuity; the Bloch eigenspace of non-singular flat bands under periodic boundary conditions is spanned by combining CLSs with extended lattice-traversing eigenstates called noncontractible loop states (NLSs)~\cite{Rhim_2019,Rhim-Yang_2021}. For finite systems, these NLSs manifest as modes that form closed loops along the system boundary and are strictly localized to it~\cite{Rhim_2019}. These so-called robust boundary modes (RBMs) cannot be disrupted through local perturbations, and serve as a manifestation of boundary effects protected by the real-space topology of the underlying lattice~\cite{Bergman2008,Rhim_2019} in contrast to the more widely established momentum-space topological protection~\cite{Hasan2010,Lu2014,Huber2016}. Besides the existence of RBMs and real-space topological phenomena, singular flat bands also generate unusual features in the energy spectrum of electrons in a magnetic field~\cite{Rhim_2020}.

To date, the exploration of singular flat-band physics has primarily been advanced using photonic lattices which can be fabricated in desired geometries via laser writing~\cite{Xia2018,Ma_2020,Xie2021,Hanafi-2022}. However, photonic lattices and their resulting band structures cannot be tuned after fabrication. By contrast, many techniques exist to tune the vibrational spectra of artificial acoustic and phononic structures through external electromagnetic or mechanical actuation~\cite{Wang2020}, enabling dynamic control of band dispersion towards and away from a flat band. In a previous theoretical and computational study, we introduced a design for a one-dimensional acoustic metamaterial based on plate resonators, with a phonon band which can be dynamically tuned to be dispersionless by applying a global tension~\cite{Karki_2021}. Our work demonstrated that dynamic tuning of a flat band can be exploited to stop and reverse a sound pulse. However, flat bands in one dimension are guaranteed to be non-singular~\cite{Rhim_2019}. Realizing singular flat bands in higher-dimensional tunable acoustic metamaterials could open up a highly adaptable platform for investigating singular flat band physics, and also provide design strategies for novel sound manipulation and processing in mechanical metamaterials.

\begin{figure}[tb]
	\noindent\includegraphics[width=\columnwidth]{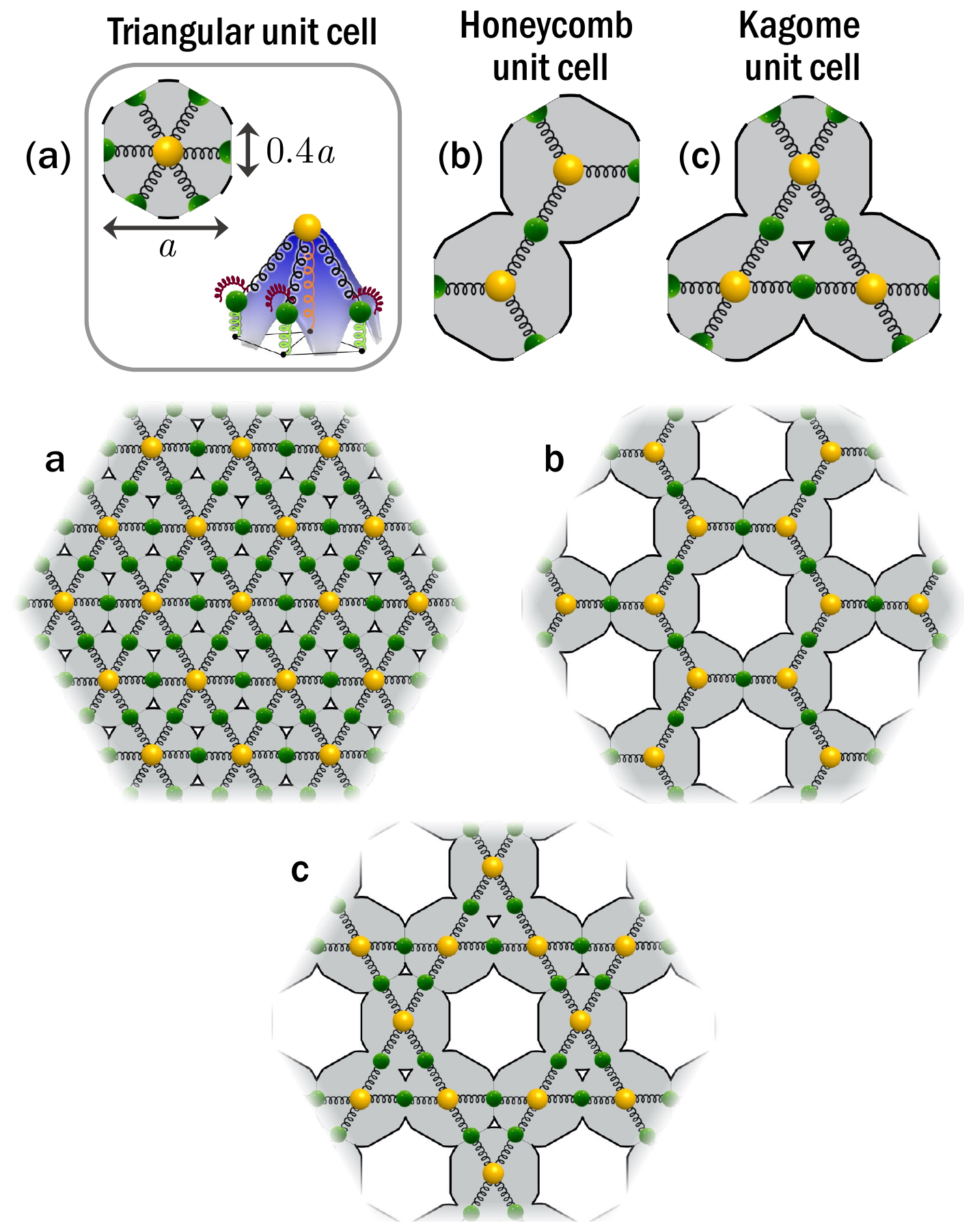}   
	\caption{   
		\label{Intro-fig}
		Unit cells of the three lattices: (a) triangular, (b) honeycomb, and (c) kagome. Discrete spring-mass models for coupled fundamental modes of the continuum thin-plate resonator metamaterial are superimposed. (a) A single resonator as a building block has a length of $a$ with the chosen junction length of $0.4a$. An example of the displacement field in the triangular unit cell is shown in the bottom panel of (a). The masses in the corresponding discrete model have vertical springs acting as anchors along with coupling springs. The bold black edges around the plates signify Dirichlet or clamped boundary conditions. On the bottom panel, figures a, b, and c show the extended version of the respective lattices.
              }
\end{figure}

As a step towards this goal, in this study we propose a class of thin-plate acoustic metamaterials in which flat bands of both singular and non-singular type can be realized. Our designs are based on the triangular lattice and its derivatives, the honeycomb and kagome lattices (\figref{Intro-fig}). The dispersion relations of the vibrational band structures depend strongly on the lattice geometry as well as the in-plane tension within the plate, a parameter which can be tuned post-fabrication using electrostatic~\cite{Cha_2018,Mei_2018} or thermally-induced~\cite{Blaikie_2019} stresses in micromechanical resonator arrays~\cite{Zande_2010}. By mapping the mechanical response of the continuum plate resonators to a discrete model of masses connected by springs~\cite{Karki_2021} (green and yellow balls in \figref{Intro-fig}), we elucidate the separate mechanisms for flat bands of different types to arise in our model, identify the conditions on system parameters that generate flat bands in the three lattices, and show that the honeycomb and kagome lattices can harbor doubly- and triply-degenerate flat bands respectively. We verify the discontinuity in the Bloch eigenfunctions of the flat band in the discrete kagome lattice model, thereby establishing the singular nature of the band. We build a boundary mode from states belonging to the singular flat band, and demonstrate its robustness against perturbations in dynamical simulations. We wrap up with a brief discussion of potential experimental platforms and of possible future directions.

\section{Theoretical framework}

\subsection{Model of coupled plate resonators} 

Our analysis uses the model of elastic plate resonator assemblies introduced in Ref.~\onlinecite{Karki_2021}. We model a continuum elastic plate (grey region in \figref{Intro-fig}) with clamped external and internal edges (solid black lines) that impose the chosen plate geometry. We study three lattice geometries that are assembled by joining identical hexagonal plates with rounded edges along prescribed boundaries to generate a triangular, honeycomb, or kagome lattice (\figref{Intro-fig} \subfiglabel{a}, \subfiglabel{b}, and \subfiglabel{c} respectively). The equation of motion for the transverse displacement field $u(x,y)$ of the plate with mass per unit area $\rho$, bending modulus $D$, and with a uniform in-plane tension $T^\prime$ is~\cite{Timoshenkotheory_1959}:
\begin{align} \label{full-pde}
	\begin{split}	
		\rho \frac{\partial^2 u}{\partial t^2} +
		D\nabla ^4 u - T^\prime\nabla^2 u = 0 \hspace{0.5cm} \text{on domain}, \\ ~~{}
		u=\nabla u=0 \hspace{0.5cm} \text{on boundary}.
	\end{split}
\end{align}
As mentioned previously, the tension is an externally-imposed prestress which can be tuned whereas the other parameters and the geometry are fixed at fabrication. 

The continuum plate equations can be non-dimensionalized by defining the dimensionless variables  $\bar{x}=x/a$, $\bar{y}=y/a$,  and $\bar{t}=t\sqrt{D/(\rho a^4)}$. In terms of these variables, \eqnref{full-pde} becomes 
\begin{align} \label{non-dim-pde}
\begin{split}	
\frac{\partial^2 u}{\partial \bar{t}^2} 
+\bar\nabla^4  u  -  T \bar\nabla^2 u = 0 \hspace{0.5cm} \text{on domain}, \\ ~~{}
u=\bar\nabla u=0 \hspace{0.5cm} \text{on boundary}. 
\end{split}
\end{align} 
For a particular boundary geometry, \eqnref{non-dim-pde} shows
that under appropriate length and time units, the system depends on a single dimensionless
parameter---the rescaled tension $\tens \equiv T^\prime a^2/D$, which serves as
the tunable physical quantity in our study.
In the remainder of this
manuscript, the bar is dropped for clarity; the variables $x$, $y$, $t$ and the gradient
operator $\nabla$ refer to the rescaled coordinates from here on.

Oscillatory solutions to \eqnref{non-dim-pde} can be written as a superposition of normal modes
$u_i(x,y)e^{-i \omega_i t}$, where the eigenfunction
$u_i(x,y)$ and oscillation frequency $\omega_i$ of the $i$th mode solve the eigenvalue problem
\begin{equation}
\label{eq:eigenvalue}
\nabla^4 u_i - T \nabla^2 u_i = \omega_i^2 u_i
\end{equation}
under the prescribed boundary conditions. We use finite-element analysis to compute normal mode eigenfunctions and frequencies (details are provided in \appref{appendixFiniteElementAnalysis}). While a true continuum system has infinitely many degrees of freedom, and thus an infinite number of normal modes per resonator, we focus in this work upon collective modes that arise from combinations of the lowest-frequency or fundamental mode of each individual resonator (schematically depicted in \subfigref{Intro-fig}{a}). Since modes of an individual resonator with higher wavenumbers are significantly higher in frequency than the fundamental mode, a system of $N$ coupled resonators will typically have its $N$ lowest normal modes composed primarily of combinations of fundamental modes on individual resonators, such as the modes shown in \subfigref{Mode-crossing}{c--d}. For an infinite periodic lattice, the eigenmodes are Bloch functions defined on a unique set of crystal momenta termed the Brillouin zone and the corresponding eigenfrequency surfaces are the bands; here, for a unit cell with $n$ resonators the lowest $n$ bands can be identified as being built primarily from the fundamental modes on individual resonators. The fundamental mode frequency $\omega_0$ for a single resonator is used as the frequency scale for mode frequencies and band structures evaluated numerically.

\subsection{Discrete model of coupled fundamental modes} 

\begin{figure}[tb]
	\noindent\includegraphics[width=\columnwidth]{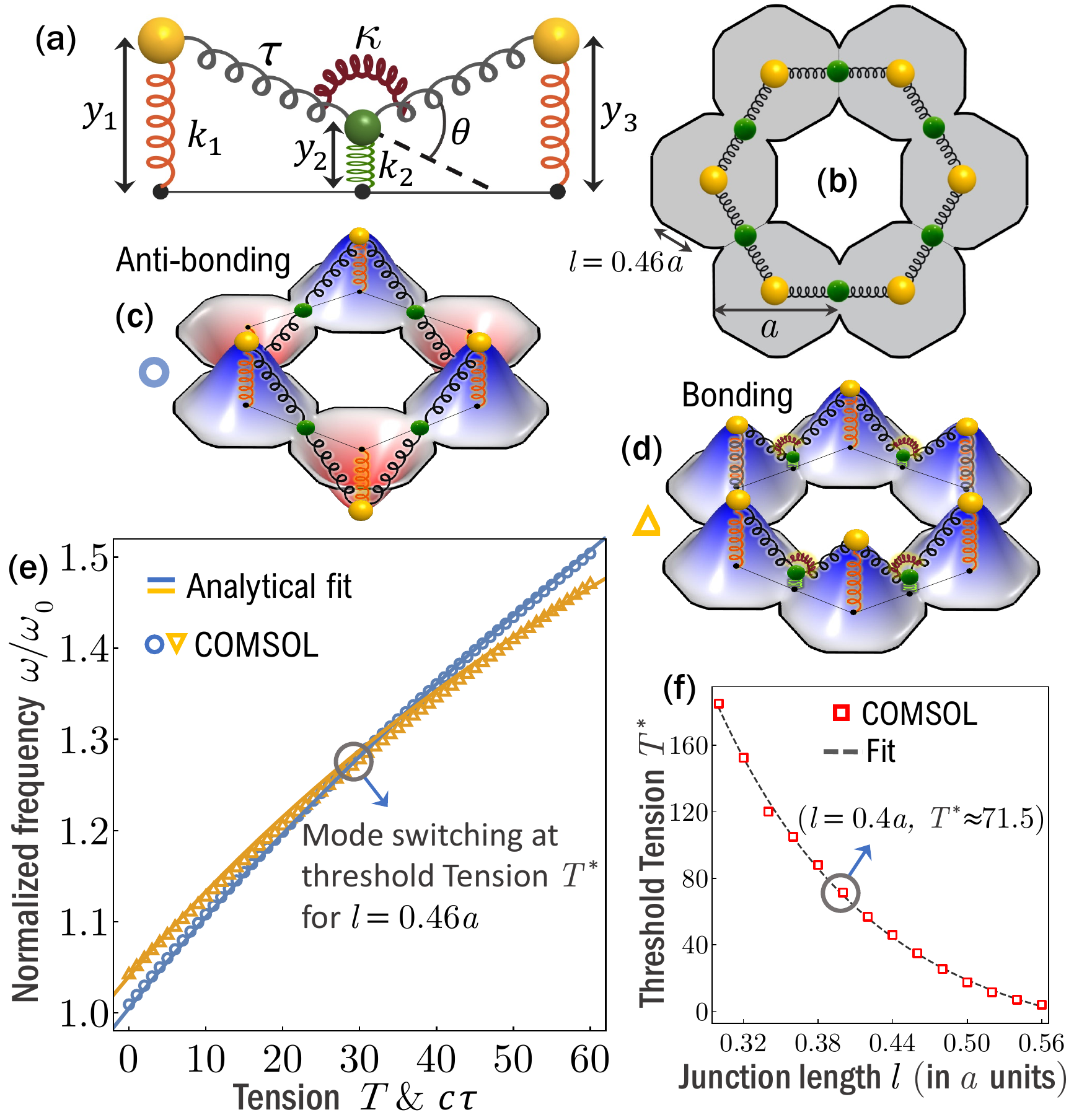}   
	\caption{   
		\label{Mode-crossing}
		Crossing between the bonding and the anti-bonding eigenmodes at threshold tension $T^*$ that can be tuned by changing the junction length in the continuum model. (a) Spring-mass discrete model describing the spring stiffnesses and the vertical displacements of the vertical springs and the angular displacement of the torsional spring. (b) Honeycomb ring with the dimensions for which the eigenmode crossing is shown in (e). The anti-bonding and the bonding modes of the 6-site honeycomb ring along with the decorated 12-site decorated honeycomb discrete model is shown in (c) and (d) respectively.  (e) The crossing between the bonding and the anti-bonding mode for the junction length $l=0.46a$ of the honeycomb ring. The modes associated with the results are shown in (c) and (d) along with the 12-site spring-mass model that gives agreeable analytical fits (solid lines). The fundamental eigenmode used for the normalizing frequency $\omega_0$ is shown in \figref{triangular-bands}(b). The change in the threshold tension $T^*$ as a function of the junction length $l$ demonstrates the geometric tunability as shown in (f). The open square markers are results from simulations and the dashed line is a fit to an exponentially decaying function $A+B\exp[-Cl]$, with $A\approx-23.10$, $B\approx2233.14$, and $C\approx7.95$.
              }
\end{figure}

The collective excitations built from couplings among fundamental modes can be described by a
simpler ``tight-binding'' description of discrete oscillators with finite-ranged
couplings, derived in Ref.~\onlinecite{Karki_2021} and summarized here. The key elements are shown as balls connected by springs in \figref{Intro-fig} (top view of different lattices) and \subfigref{Mode-crossing}{a} (side view of a minimal unit comprising a pair of coupled resonators). The primary degrees of freedom are the fundamental modes of individual resonators, each of which is modeled as a harmonic oscillator of mass $m$ confined to the vertical direction (yellow balls), with spring stiffness $\tilde{k}_1$ (red spring). To correctly model the effect of variable tension $T$ on the coupling between fundamental modes of adjacent resonators, we introduce an additional, secondary degree of freedom, also a ball of mass $m$,  which encodes the vertical displacement of the plate at the junction between two resonators (green ball and spring) with spring constant $\tilde{k}_2$. Since the junction between resonators is narrower than the resonator diameter, the secondary degree of freedom has a stiffer spring 
$\tilde{k}_2 > \tilde{k}_1$. The junction mass is coupled to the fundamental mode masses by harmonic springs (dark grey) which are prestressed with a tensile force $\tilde{\tau}$, and a torsional spring (red) which induces the two tensile springs to be collinear. If $y_1$ and $y_3$ correspond to the vertical displacements of the resonator degrees of freedom and $y_2$ to that of the junction, the effect of the tensile and torsional springs is captured in potential energy contributions  $U_s = \tilde{\tau}[(y_1-y_2)^2+(y_2-y_3)^2]/a$ and $U_b = \tilde{\kappa}(1-\cos\theta) \approx 2\tilde{\kappa} (y_1 -
2y_2+y_3)^2/a^2$ respectively.

Newton's equations of motion for the coupled spring-mass system can be written as
\begin{equation}
m\frac{d^2\mathbf{y}}{d\tilde{t}^2}+\tilde{\mathbf{K}}\mathbf{y}=0,
\end{equation} 
where $\mathbf{y} = \{y_1,y_2,...\}$ is the vector of vertical displacements,
and $\tilde{\mathbf{K}}$ is the stiffness matrix whose entries are obtained by taking derivatives of the total potential energy in the on-site, tensed, and torsional springs. For the three-site model of a coupled resonator pair shown in \subfigref{Mode-crossing}{a}, the stiffness matrix reads
\begin{equation}
\tilde{\mathbf{K}} = {\begin{pmatrix} \tilde{k}_1+\frac{\tilde{\tau}}{\ell}+\frac{\tilde{\kappa}}{\ell^2} & -\frac{\tilde{\tau}}{\ell}-\frac{2\tilde{\kappa}}{\ell^2} & \frac{\tilde{\kappa}}{\ell^2} \\ -\frac{\tilde{\tau}}{\ell}-\frac{2\tilde{\kappa}}{\ell^2} & \tilde{k}_2+\frac{2\tilde{\tau}}{\ell}+\frac{4\tilde{\kappa}}{\ell^2} & -\frac{\tilde{\tau}}{\ell}-\frac{2\tilde{\kappa}}{\ell^2} \\ \frac{\tilde{\kappa}}{\ell^2} & -\frac{\tilde{\tau}}{\ell}-\frac{2\tilde{\kappa}}{\ell^2} & \tilde{k}_1+\frac{\tilde{\tau}}{\ell}+\frac{\tilde{\kappa}}{\ell^2} \end{pmatrix}} 
\label{DynamicalMatrix-3M-full}
,
\end{equation}
where $\ell$ is the horizontal spacing between the masses.
To build a discrete model with
dimensionless parameters that can be related to the continuum system, we rescale time and displacements by $\omega_0^{-1}$ and $a$ respectively. The distance
between primary degrees of freedom $y_i$ and $y_{i+2}$ is also set to be
$a$, so that $\ell = a/2$. In terms of the rescaled time $t=\omega_0\tilde{t}$, spring stiffnesses
$k_i=\tilde{k_i}/m\omega_0^2$, the tension $\tau=2\tilde{\tau}/am\omega_0^2$,
and the torsional stiffness $\kappa=4\tilde{\kappa}/a^2m\omega_0^2$, \eqnref{DynamicalMatrix-3M-full} reduces to
\begin{equation}
\mathbf{K} =  \left(
  \begin{array}{ccc}
    k_1 +\tau+\kappa  & -\alpha  & \kappa   \\
    -\alpha  & k_2+2\tau+4\kappa  & -\alpha   \\
    \kappa & -\alpha  & k_1+\tau+\kappa  \\
  \end{array}
\right)
\label{DynamicalMatrixTriple},
\end{equation}
where $\alpha=\tau+2\kappa$ is the net nearest-neighbor coupling strength. The torsional springs generate next-nearest-neighbor couplings with strength $\kappa$. The non-dimensionalized equation of motion is
\begin{equation}
\frac{d^2\mathbf{y}}{dt^2}+\mathbf{K} \mathbf{y}=0,
\end{equation}
and the normal modes $\mathbf{u}_i e^{-i \omega_i t}$ of the discrete system are obtained from solutions of the eigenvalue equation
\begin{equation}
  \label{eq:discreteeigen}
  \mathbf{K} \mathbf{u}_i = \omega_i^2 \mathbf{u}_i.
\end{equation}
For a continuum system with a given set of physical parameters, the parameters of the corresponding discrete model are obtained by matching the normal mode frequencies of the modes arising from the fundamental resonator excitations of the continuum system~\cite{Karki_2021}. The structure and symmetries of the discrete stiffness matrix provide insights into the spectral features of the continuum system and their relationship to the geometry and tunable parameters.

\subsection{Tunable mode-crossing mechanism}

The stiffness matrix for arbitrary networks of coupled resonators can be built by overlapping blocks of the form of \eqnref{DynamicalMatrixTriple}. As an example of matching discrete and continuum models which elucidates the mode crossing mechanism that generates tunable flat bands, we consider a six-resonator ring, \subfigref{Mode-crossing}{b}, which is a recurring motif in the honeycomb and kagome lattices. We expect the continuum eigenmodes with the lowest six eigenfrequencies to derive primarily from the fundamental modes of individual resonator. At $T=0$, the lowest eigenmode is shown in \subfigref{Mode-crossing}{c} and can be interpreted as a combination of fundamental modes in an ``anti-bonding'' configuration with displacements on  adjacent resonators bearing opposing signs. The sixth eigenmode is a ``bonding'' configuration, shown in \subfigref{Mode-crossing}{d}. The intermediate modes involve other combinations of the six fundamental modes.

Upon increasing the tension, the anti-bonding and bonding modes approach each other in frequency, as shown in \subfigref{Mode-crossing}{e}. At a threshold tension $T^*$, the frequencies of the two modes become degenerate with each other and with the four intermediate modes. When the tension is increased further, the bonding mode becomes the lowest in frequency, as dictated by the maximum principle when the Laplacian operator dominates the plate equation, \eqnref{non-dim-pde}~\cite{Sweers_2001,Brown_1999,Karki_2021}. The degeneracy in the coupled modes at $T^*$ demonstrates one mechanism for generating flat bands in a periodic 2D array of resonators.

The balance of bending and stretching energy that gives rise to the mode crossing is made evident in the corresponding discrete model. The discretized six-resonator ring has twelve harmonic degrees of freedom, whose dynamics are dictated by the $12\times 12$ stiffness matrix
\begin{equation}
\mathbf{K} =  \left(
							\begin{array}{cccccccccccc}
								\phi_1  & -\alpha  & \kappa  & \dots & \dots & \kappa  & -\alpha  \\
								-\alpha  & \phi_2  & -\alpha  & \dots & \dots & \dots & 0  \\
								\kappa & -\alpha  & \phi_1  & -\alpha  & \kappa & \dots & 0   \\
								0 & 0 & -\alpha & \phi_2 & -\alpha & \dots & 0 \\
								\vdots  & \ddots  & \ddots & \ddots & \ddots  & \vdots & \vdots \\
								\kappa  & \dots & \dots & \kappa  & -\alpha  & \phi_1  & -\alpha  \\
								-\alpha  &  \dots & \dots & \dots & \dots & -\alpha  & \phi_2 \\
							\end{array}
						\right)
\label{DynamicalMatrix-3M},
\end{equation}
where $\phi_1=k_1+2\tau+2\kappa$ is the on-site net stiffness of the primary mass and $\phi_2=k_2+2\tau+4\kappa$ is the on-site net stiffness of the junction mass. Out of the twelve eigenvectors of the discrete model, displacements corresponding to the  anti-bonding and bonding modes can be readily identified (masses and springs in \subfigref{Mode-crossing}{c--d}) and the corresponding eigenfrequencies are respectively
\begin{align} 
\begin{split}
\omega_1^2 &=k_1+2\tau, \\
\omega_2^2 &=\frac{k_1+k_2+4 (\tau+2 \kappa)-\sqrt{\text{U}}}{2},	
\label{3M-eigenvalues}
\end{split}
\end{align}
where $\text{U}=(k_1-k_2)^2+ 16(\tau + 2\kappa)^2$. We use these expressions to fit the discrete model to the eigenfrequencies of the continuum mode. Specifically, we treated $k_1$, $k_2$, and $\kappa$ as constants across all tension values, and assumed a linear relationship $T = c\tau$, where $c$ is a constant parameter. 
Given the exact frequencies of the discrete model, \eqnref{3M-eigenvalues}, the parameter value $k_1=1.01$ is fixed by equating it to the square of antibonding mode frequency from the continuum model at $T = \tau = 0$. The complete relationship between frequency and prestress (tension) for this mode is then quantitatively recovered by setting $c = 95$. Having set these two parameters, the remaining parameters $k_2=10$ and $\kappa=0.018$ were fixed by fitting the analytical form for $\omega_2$ from \eqnref{3M-eigenvalues} to the bonding mode frequency curve from the continuum model. Through this procedure, we found that the discrete model with four fit parameters quantitatively captures the dependence of normal mode frequencies on the varying tension measured in the continuum model (compare symbols to solid lines in \subfigref{Mode-crossing}{e}).

The lowest six modes in the discrete model become degenerate when the coupling constants satisfy the relation
$$\tau = \sqrt{(2\kappa)^2 + \kappa(k_2-k_1)}-2\kappa.$$
For fixed values of the stiffnesses $k_1$, $k_2$ and $\kappa$, a tension can always be found to generate a mode crossing provided $k_2 > k_1$. This constraint on effective parameters tends to be satisfied by coupling regions that are narrower than the typical extent of an individual resonator. Physically, tuning the relative strength of tension and bending in the system has the effect of flipping the sign of the effective interaction between the primary masses in a tight-binding description; our model realizes a mechanical analogue of a method proposed for sign control of coupling terms in tight-binding models in Ref.~\onlinecite{Keil2016}, which also uses a third degree of freedom between the primary sites to mediate the sign change. As a design principle, it useful to consider the fundamental mode on each resonator as a degree of freedom in a tight-binding model. Each degree of freedom is coupled to degrees of freedom on neighboring resonators. The strength and sign of the effective coupling are controlled by the global tension on the plate. At the degeneracy point, the coupling strength is effectively zero as bonding and anti-bonding pairs have equal vibration frequency; as a consequence, we expect a degeneracy among all fundamendal modes present in the assembly. This mechanism, when applied to an infinite lattice of coupled resonators, suggests that flat bands will generically be present at special values of the in-plane tension. 

The threshold tension $T^*$ occurs at a particular value for a given geometry, and also sets the frequency of the degeneracy. Once the geometry is fixed, the system can be moved towards or away from the degeneracy point, but the frequency of the degenerate modes cannot be changed. To tune the threshold tension and the degenerate frequency for a particular resonator geometry, we can adjust the junction length $l$ relative to the resonator size $a$. In \subfigref{Mode-crossing}{f}, we show that the threshold tension value of the mode crossing in continuum simulations can be varied significantly by changing the junction length. 
Therefore, changing tension allows dynamical tuning and changing geometry allows the threshold tension tuning, providing two types of tunability in the system. Whereas we will focus on the effect of changing tension for fixed lattice geometries in the next section, we mention here that the junction geometry could also be chosen in advance to target particular tension ranges or frequency values for the flat band. 

\section{Results}

\subsection{Band structures} 

We now investigate the consequences of the mode-crossing mechanism for infinite periodic lattices of resonators. We restrict ourselves to lattices based on the triangular Bravais lattice, namely the triangular, honeycomb, and kagome lattices (\figref{Intro-fig}) as both non-singular and singular flat bands are observed within this set of lattices. 

We first define variables which will be used among all lattices. The three lattice vectors defining the triangular Bravais lattice are  $\bm{a}_1=(a,0)$, $\bm{a}_2=(a/2,\sqrt{3}a/2)$, and $\bm{a}_3=\bm{a}_2-\bm{a}_1=(-a/2,\sqrt{3}a/2)$, and $\bm{q}=(q_x,q_y)$ is the crystal momentum. The following variables allow us to present stiffness matrices for the discrete models succintly:
$\bm{\upgamma}_i=e^{i \bm{q}.\bm{a}_i}$ and $\bm{\upzeta}_i=1+\bm{\upgamma}_i$, with $\bm{\upgamma}^*$ and $\bm{\upzeta}^*$ denoting their respective complex conjugates.

\begin{figure}[tb]
	\noindent\includegraphics[width=\columnwidth]{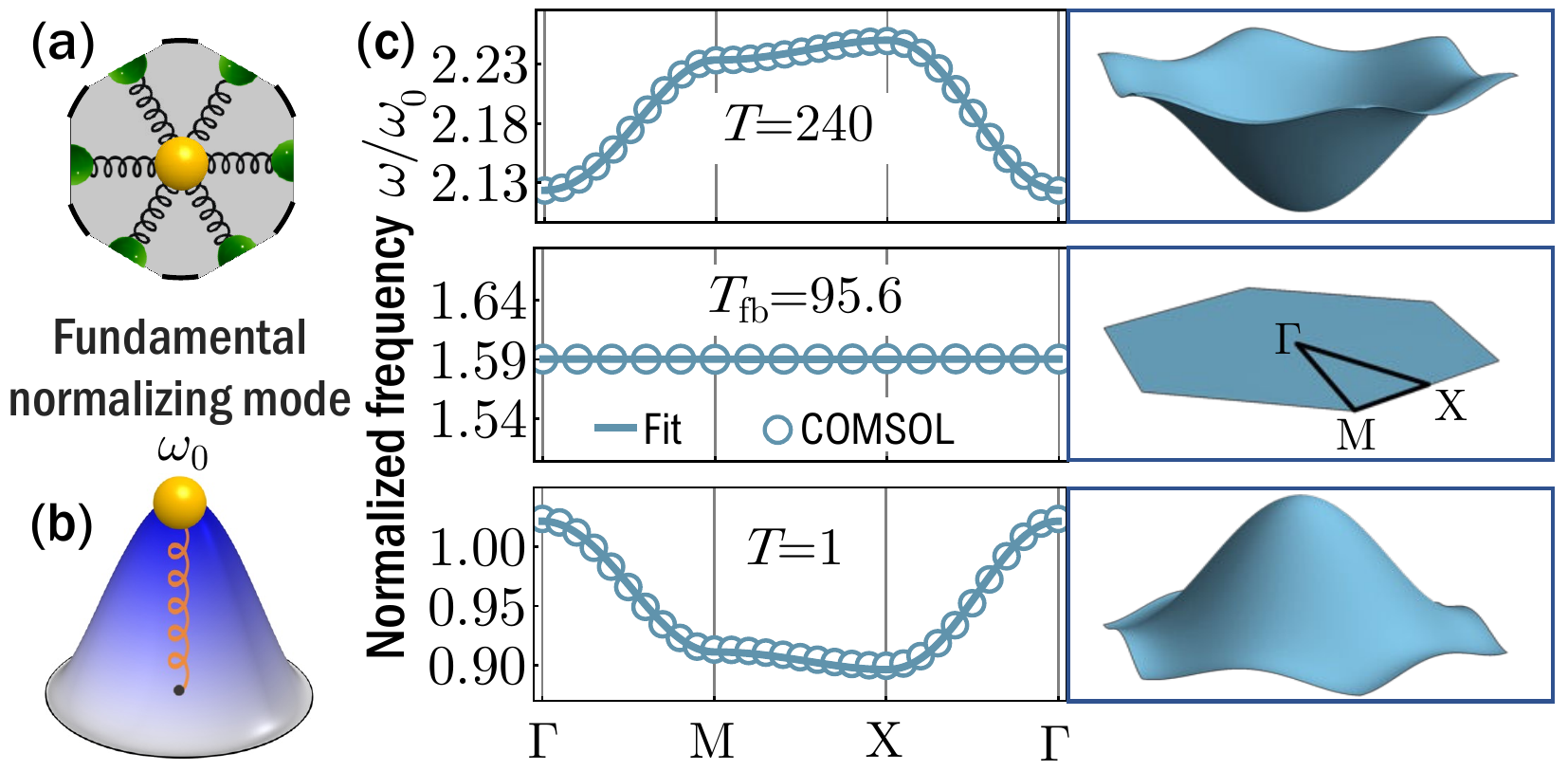}   
	\caption{   
		\label{triangular-bands}
		Band structure of the triangular lattice for three different tension values. (a) Unit cell with the corresponding discrete model
		superimposed. (b) The fundamental mode of a circular clamped plate ($r\approx0.54a$) with frequency $\omega_0\approx34\sqrt{D\rho/a^4}$ for $T=0$ and 
		$D=1$. The frequency $\omega_0$ is used as the normalizing frequency throughout this study.
		(c) Dispersion relation of the lowest band from finite-element computations for the continuum model (symbols) and analytical evaluation of the discrete model with parameters obtained by a fit to the continuum model (solid curves/surfaces). Values are shown along line segments connecting the high-symmetry points on the left, and for the entire hexagonal Brillouin zone on the right. The high-symmetry points $\Gamma$, $\text{M}$, $\text{X}$ and the segments connecting them are shown on the surface plot of the band at $T = T_\text{fb}$ for which the band is completely flat. 
		Discrete model parameters $\kappa=0.19$ and $k_2=2.62$ are held constant for all tension values. Fit parameters $\tau$ and $k_1$ for different tensions $T$ are: $T=1$
		$\rightarrow$ $\{\tau,k_1\}\approx\{0.39, 0.36\}$, $T=95.6\equiv T_\text{fb}$
		$\rightarrow$ $\{\tau,k_1\}\approx\{0.59, 1.06\}$, and $T=240$
		$\rightarrow$ $\{\tau,k_1\}\approx\{0.73, 1.78\}$.
              }
\end{figure}

\subsubsection{Triangular lattice}
In the thin-plate continuum metamaterial system, the triangular lattice is a Bravais lattice and can be transformed into either the honeycomb lattice or the kagome lattice 
by removing subsets of sites as seen in \figref{Intro-fig}. The Brillouin zone of the triangular lattice is a hexagonal region in momentum space, and the bands have three high-symmetry points marked as $\Gamma$, $\text{M}$ and $\text{X}$, see labels in \subfigref{triangular-bands}{c}  and \subfigref{Kagome-bands}{b}. The unit cell in the continuum model has a single resonator (\subfigref{triangular-bands}{a}), and therefore the lowest-frequency band is associated with the fundamental modes on individual resonators, as we verify by examining the numerically computed eigenfunctions for states from the lowest band. The continuum dispersion relations were calculated by sampling 31 evenly-spaced points along the segments connecting the high-symmetry points in the Brillouin zone (symbols in \subfigref{triangular-bands}{c}). 

At low tension ($T = 1$ in \subfigref{triangular-bands}{c}), the $\mathbf{q}=0$ mode has the largest frequency in the band, consistent with the ``bonding'' configuration being of higher frequency than the ``anti-bonding'' configuration because the displacements on all resonators have the same phase at $\mathbf{q}=0$~\cite{Karki_2021}. At high tension ($T = 240$), the relative frequencies at the band center and the band edges have flipped, and $\mathbf{q}=0$ now bears the lowest frequency, consistent with a flip in the order of bonding and anti-bonding configurations. At a specific value of the tension $T = T_\text{fb} = 95.6$, the band is completely flat across all sampled points, showing that the degeneracy mechanism observed for the six-resonator ring gives rise to a flat band at a fine-tuned parameter value. Since the lowest band is isolated from higher bands and does not touch any other band, the flat band at $T_\text{fb}$ is a non-singular band as a singularity can only occur at a band touching point.

While the continuum model is in the triangular lattice form, the discrete model is a decorated triangular lattice since an extra lattice point is 
required to model the junction between each resonator as shown in \subfigref{Intro-fig}{a}.
The unit cell of the discrete model has four degrees of freedom, resulting in four frequency bands.
The $4\times4$ matrix describing the Bloch bands of the periodic lattice in the discrete model is
\begin{equation} \label{tri-model}
	\mathcal{\mathbf{K}}_{\text{T}}(\bm{q})=\left(
	\begin{array}{cccc}
		V_1- 2 \kappa \sum\cos \bm{q}.\bm{a}_i & \alpha\bm{\upzeta}_3^* & \alpha\bm{\upzeta}_2^*  & \alpha\bm{\upzeta}_1^* \\
		\alpha\bm{\upzeta}_3 & V_2 & 0 & 0 \\
		\alpha\bm{\upzeta}_2 & 0 & V_2 & 0 \\
		\alpha\bm{\upzeta}_1  & 0 & 0 & V_2 \\
	\end{array}
	\right),
\end{equation}
where $V_1=k_1+4 \tau+6 \kappa$ is the on-site stiffness for the primary degree of freedom and $V_2=k_2+2 \tau+4 \kappa$ is the on-site stiffness for the secondary degrees of freedom. The frequency bands are then solved via
$|\mathcal{\mathbf{K}}(\bm{q})-\omega(\bm{q})^2 \mathbf{I}|=0$, giving rise to four bands. When $k_2 > k_1$, the lowest band includes eigenvectors whose displacements are significantly larger for the primary degree of freedom than for the secondary degrees of freedom, consistent with our assignment of the resonator and junction displacements with the primary and secondary oscillators respectively.

To fix the discrete model parameters, we used a fit of the analytical form for the lowest eigenfrequency from the discrete model to the numeric non-dimensionalized frequencies $\omega/\omega_0$ for the lowest band from the continuum model along the line segments connecting the symmetry points $\Gamma$, $\text{M}$ and $\text{K}$. The numerically obtained frequencies from the continuum model (symbols) are compared to the fitted analytical solution for the lowest eigenfrequency of \eqnref{tri-model} (curves) in the left column of \subfigref{triangular-bands}{c}, showing good agreement at all three tension values. We were able to obtain a quantitative fit across the Brillouin zone by keeping the parameters $\kappa$ and $k_2$ fixed across all values of $T$, and allowing both $k_1$ and $\tau$ to vary at each $T$ value (see caption to \figref{triangular-bands} for fit parameter values). The variation in both the tensile force and the on-site harmonic stiffness in the discrete model reflects the fact that changing the continuum tension modifies the fundamental mode frequency on each resonator, and also influences the coupling between modes on adjacent resonators. 

In the discrete model, the lowest band becomes completely flat (i.e. the eigenfrequency has no $\mathbf{q}$-dependence) at a parameter value $\tau$ that is related to the other discrete parameter values via
\begin{equation} \label{eq:triangular-flat}
\tau = \sqrt{\left(3\kappa\right)^2+\kappa(k_2-k_1)}-3\kappa.
\end{equation}
If we treat the stiffnesses $k_1$, $k_2$ and $\kappa$ as fixed and the spring tension $\tau$ as a variable parameter, we can find a real value of $\tau$ that generates a flat band as long as $k_2>k_1$. This requirement tends to be satisfied for fits of our discrete model to continuum data across different geometries and tension values, and reflects the physical constraints to the model for narrow junctions: displacements at the junctions cost more elastic energy than the same displacements at the center of the resonators, which translates to a higher stiffness for the junction degrees of freedom in the discrete model. 

\subsubsection{Honeycomb lattice}

\begin{figure}[tb]
	\noindent\includegraphics[width=\columnwidth]{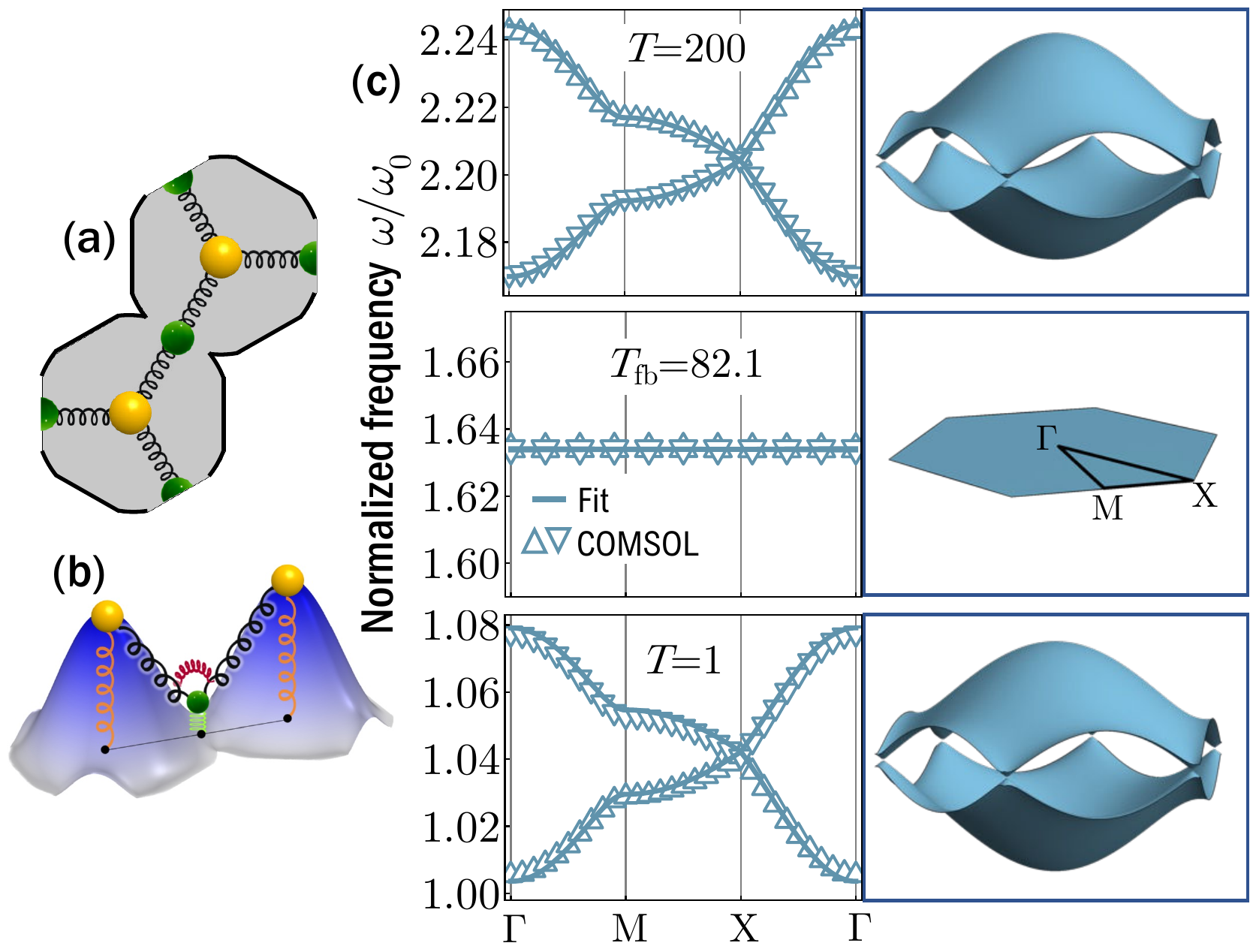}   
	\caption{   
		\label{honeycomb-bands}
		Band structures of the honeycomb lattice for three different tension values. (a) Unit cell with the corresponding discrete model
		superimposed. (b) An example eigenmode of the unit cell at the symmetry point M of Brillouin zone is shown along with the corresponding discrete model. 
		The discrete model also has a torsional spring which is activated due to the bending at the junction replicating the bending stiffness of the continuum model. (c) Dispersion relations of the lowest two bands from the continuum (symbols) and discrete (curves/surfaces) models at three different tension values. Values are shown along line segments connecting the high-symmetry points on the left, and as a surface plot over the entire hexagonal Brillouin zone on the right. Note that there are two bands in all three surface plots, but they exactly coincide for $T=T_\text{fb}$. 
		Parameters $\kappa=0.19$ and $k_2=2.62$ are held constant. Fit parameters $\tau$ and $k_1$ for different tensions $T$ are: $T=1$
		$\rightarrow$ $\{\tau,k_1\}\approx\{0.39, 0.74\}$, $T=82.1\equiv T_\text{fb}$
		$\rightarrow$ $\{\tau,k_1\}\approx\{0.59, 1.28\}$, and $T=200$
		$\rightarrow$ $\{\tau,k_1\}\approx\{0.68, 1.91\}$.}
\end{figure}

The honeycomb lattice is built by decorating the triangular Bravais lattice with a two-resonator unit shell shown in \subfigref{honeycomb-bands}{a}. Consequently, the lowest two bands in frequency are associated with the fundamental resonator modes. We expect a dispersion similar to that of the tight-binding honeycomb lattice, used as a basic model for electrons in graphene~\cite{CastroNeto2009}: the two bands touch each other at the six corners of the Brillouin zone and exhibit a linear dispersion in the vicinity of the band touchings (\subfigref{honeycomb-bands}{c}). At low tension, the effective coupling between fundamental modes is negative, and the lowest-frequency mode of the lowest band at $\mathbf{q}=0$ is an antibonding state where each sublattice of the honeycomb lattice shares in-phase displacements of the fundamental modes, but the two sublattices have displacements of opposite sign compared to each other. Upon increasing the tension, we cross through a point $T=T_\text{fb}=82.1$ at which the effective coupling of the fundamental modes becomes zero. At this point, the two bands become flat and degenerate. As the tension is increased further, the bands again separate, but the lowest band is now associated with a bonding configuration with all displacements in-phase at $\mathbf{q}=0$, analogous to a tight-binding model with positive couplings. 

The discretized version of the resonator lattice is a decorated honeycomb lattice as shown in \subfigref{honeycomb-bands}{a}.
The $5\times5$ stiffness matrix encapsulating the coupled discrete degrees of freedom in the decorated honeycomb lattice is
\begin{equation}
	\mathcal{\mathbf{K}}_{\text{H}}(\bm{q})=\left(
	\arraycolsep=1.5pt
	\begin{array}{ccccc}
		V_{\text{H}1} & \alpha  & -\kappa\left(\bm{\upzeta}_2^* +\bm{\upgamma}_3^*\right) & \alpha \bm{\upgamma}_2^* & \alpha \bm{\upgamma}_3^*\\
		\alpha  & V_2 & \alpha  & 0 & 0 \\
		-\kappa\left(\bm{\upzeta}_2 +\bm{\upgamma}_3\right) & \alpha  & V_{\text{H}1} & \alpha  &
		\alpha  \\
		\alpha \bm{\upgamma}_2 & 0 & \alpha  &  V_2 & 0 \\
		\alpha \bm{\upgamma}_3 & 0 & \alpha  & 0 & V_2 \\
	\end{array}
	\right),
\end{equation}
where, $V_{\text{H}1}=k_1+3\tau+3 \kappa$ is the on-site term for the primary degrees of freedom. When $k_2 > k_1$, the two lowest-frequency bands are well-separated from the three higher bands and represent the modes for which the primary degrees of freedom have much larger displacements than the secondary degrees of freedom. The analytical expressions for these two bands (solid curves in \subfigref{honeycomb-bands}{c}) are fit to the numerically-computed frequencies from the continuum model (symbols) for points along the line segments connecting the high-symmetry points in the Brillouin zone. 
In the discrete model, two degenerate flat bands occur when the parameter $\tau$ satisfies
\begin{equation} \label{eq:honeycomb-flat}
\tau = \sqrt{\left(5\kappa/2\right)^2+\kappa(k_2-k_1)}-5\kappa/2.
\end{equation}
The change in lattice topology gives rise to a different expression for flat bands compared to that for the triangular lattice, \eqnref{eq:triangular-flat}. As with the triangular lattice, a real value of $\tau$ can be found which satisfies the above condition as long as $k_2>k_1$.

\subsubsection{Kagome lattice}

\begin{figure}[tb]
	\noindent\includegraphics[width=\columnwidth]{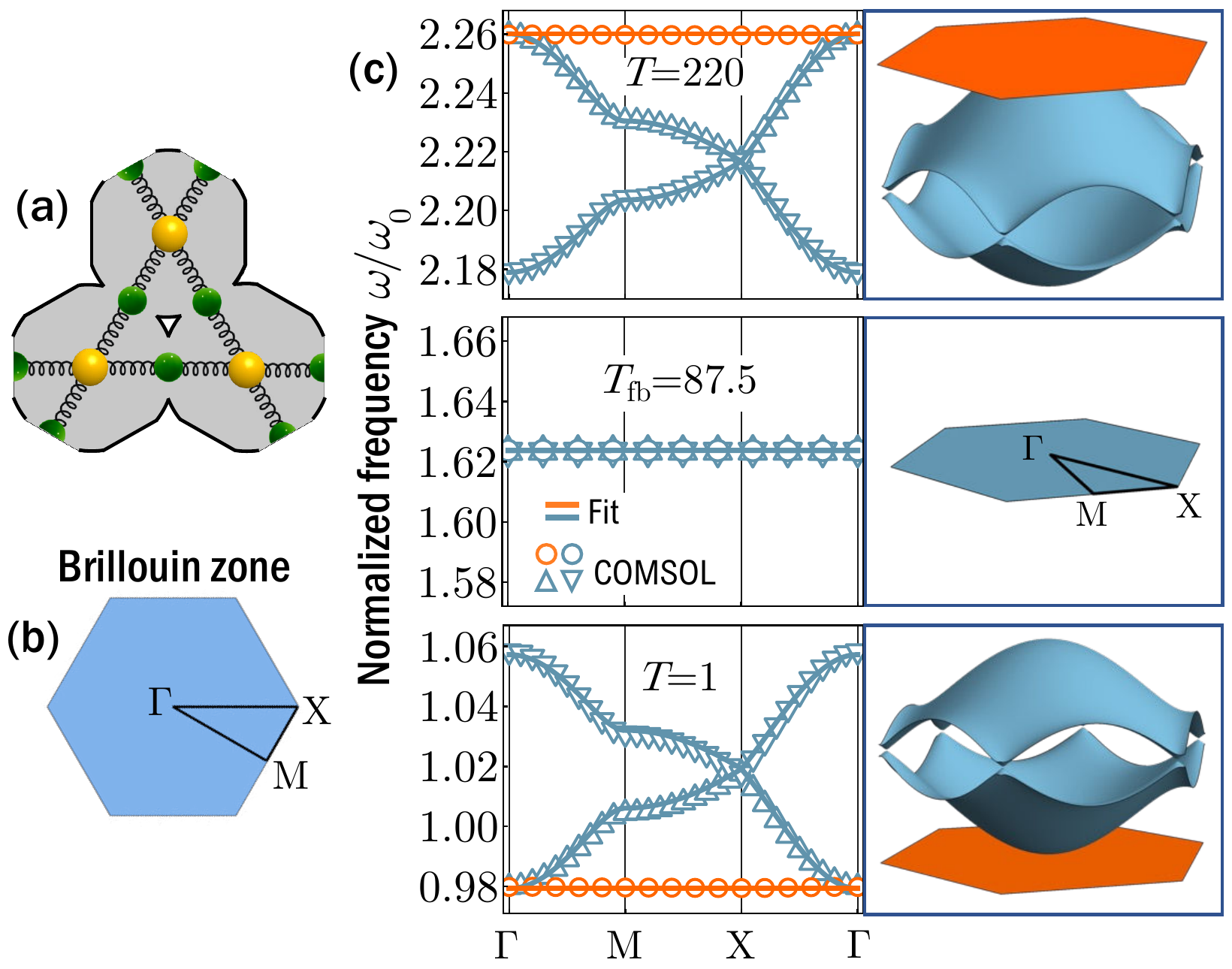}   
	\caption{   
		\label{Kagome-bands}
		Band structures of the kagome lattice for three different tension values. (a) Unit cell with the corresponding discrete model
		superimposed. (b) Brillouin zone of the kagome lattice, with high-symmetry points marked. (c) Band structures from the continuum model (symbols) and discrete model (solid curves/surfaces) show that there is always one band that 
		is flat (circular red open markers) for all values of tension. At the threshold tension of the crossing point between the bonding and the anti-bonding eigenmode, all the three bands become degenerate and flat. 
		Parameters $\kappa=0.19$ and $k_2=2.62$ are held constant. Fit parameters $\tau$ and $k_1$ for different tensions $T$ are: $T=1$
		$\rightarrow$ $\{\tau,k_1\}\approx\{0.39, 0.48\}$, $T=87.5\equiv T_\text{fb}$
		$\rightarrow$ $\{\tau,k_1\}\approx\{0.59, 1.08\}$, and $T=220$
		$\rightarrow$ $\{\tau,k_1\}\approx\{0.71, 1.79\}$.}
\end{figure}

The flat bands in the triangular and honeycomb lattices require fine-tuning of the global in-plane tension on the resonator system: the bands in question generically have a nonuniform dispersion relation, and become dispersion-free only at a special tension value for a particular geometry.  This behavior is anticipated by the tight-binding dispersion relations of the same lattices, which display flat bands only when the nearest-neighbor coupling is set to zero. Such flat bands do not arise from the real-space lattice topology and are non-singular. By contrast, the kagome lattice is an example of a lattice which, in a tight-binding description, generically exhibits a flat band due to its lattice topology. The flat band is singular by virtue of a discontinuity in its Bloch eigenstates at a quadratic band touching with its neighboring band~\cite{Bergman2008, Rhim_2019}. 

To replicate this mechanism in our mechanical system, we built a kagome lattice of resonators using a three-site unit cell shown in \subfigref{Kagome-bands}{a}; the bands of interest are the three lowest-frequency band. Consistent with the tight-binding expectation, we find a flat band guaranteed by the lattice topology at all tension values in the continuum system (highlighted in orange in \subfigref{Kagome-bands}{c}). The flat band touches the adjacent dispersive band at $\mathbf{q}=0$, whereas the third band and second band touch at the six corners of the Brillouin zone. The band structures at low and high tension are consistent with that of a tight-binding kagome lattice model with a change in sign of the hopping term between $T=1$ and $T=220$, so that the flat band shifts from the bottom to the top in order of frequencies. At the point of the sign change, the effective coupling of the three primary degrees of freedom becomes zero, and we observe three degenerate flat bands at $T=T_\text{fb} = 87.5$.

The kagome lattice in the continuum model corresponds 
to a decorated kagome lattice in the discrete model with three primary and six secondary degrees of freedom in the unit cell as shown in \subfigref{Kagome-bands}{a}.
The corresponding $9 \times 9$ stiffness matrix is
\begin{equation}
	\mathcal{\mathbf{K}}_{\text{K}}(\bm{q})=\left(
	\arraycolsep=1.5pt
	\begin{array}{ccccccccc}
		V_1 & \alpha  &  -\kappa\bm{\upzeta}_2^* & 0 & \alpha \bm{\upgamma}_2^* & 0 & \alpha \bm{\upgamma}_1^*& -\kappa \bm{\upzeta}_1^* & \alpha  \\
		\alpha  &V_2 & \alpha  & 0 & 0 & 0 & 0 & 0 & 0 \\
		-\kappa\bm{\upzeta}_2& \alpha  & V_1 & \alpha  & \alpha  & \alpha  & 0 & -\kappa\bm{\upzeta}_3 & 0 \\
		0 & 0 & \alpha  & V_2 & 0 & 0 & 0 & \alpha \bm{\upgamma}_3 & 0 \\
		\alpha \bm{\upgamma}_2  & 0 & \alpha  & 0 &V_2 & 0 & 0 & 0 & 0 \\
		0 & 0 & \alpha  & 0 & 0 & V_2 & 0 & \alpha  & 0 \\
		\alpha \bm{\upgamma}_1 & 0 & 0 & 0 & 0 & 0 & V_2 & \alpha  & 0 \\
		-\kappa\bm{\upzeta}_1 & 0 & -\kappa\bm{\upzeta}_3^* & \alpha \bm{\upgamma}_3^*  & 0 & \alpha  & \alpha  & V_1 & \alpha  \\
		\alpha  & 0 & 0 & 0 & 0 & 0 & 0 & \alpha  & V_2 \\
	\end{array}
	\right),
\end{equation}
where the on-site terms $V_1$ and $V_2$ for the primary and secondary sites are the same as for the triangular lattice. The lowest three bands of the discrete model display the same behavior as of the continuum model: a flat band exists at all tension values, and touches a dispersive band at $\mathbf{q}=0$. At low values of $\tau$, the flat band has the lowest frequency; at high values of $\tau$, the order of the lowest three bands is inverted and the flat band has the highest frequency of the three bands (albeit still lower than the frequencies of the other six bands in the discrete model). The threshold separating these two situations occurs when
\begin{equation} \label{eq:kagome-flat}
\tau = \sqrt{7\kappa^2+\kappa(k_2-k_1)}-3\kappa.
\end{equation}
At this value of the tensile force in the discrete model, the three bands become degenerate and dispersion-free, giving rise to a triply-degenerate flat band. The threshold $\tau$ value is always real when $k_2 > k_1$. However, it can be negative if $k_2 < k_1 + 2\kappa$. Note that negative values of the tensile force $\tau$ in the coupling springs of the discrete model are physically allowed, just as negative values of the tension $T$ are allowed in the continuum model.

\subsection{Flat band singularity and Hilbert-Schmidt distance} 

The band structure of the kagome lattice is distinct from that of the triangular and honeycomb lattices in exhibiting a degeneracy  between the flat band and the dispersive band that is nearest in frequency at the crystal momentum $\bm{q}=0$. The adjacent band has a quadratic dispersion relation near the touching point, and the degeneracy is termed a  quadratic band touching (QBT). QBTs of flat bands can be classified as singular or non-singular depending on the presence or absence of a discontinuity in the Bloch eigenfunctions of each band at the touching point, which cannot be removed through a local gauge transformation~\cite{Rhim_2019}.

\begin{figure*}
  \centering
  \includegraphics{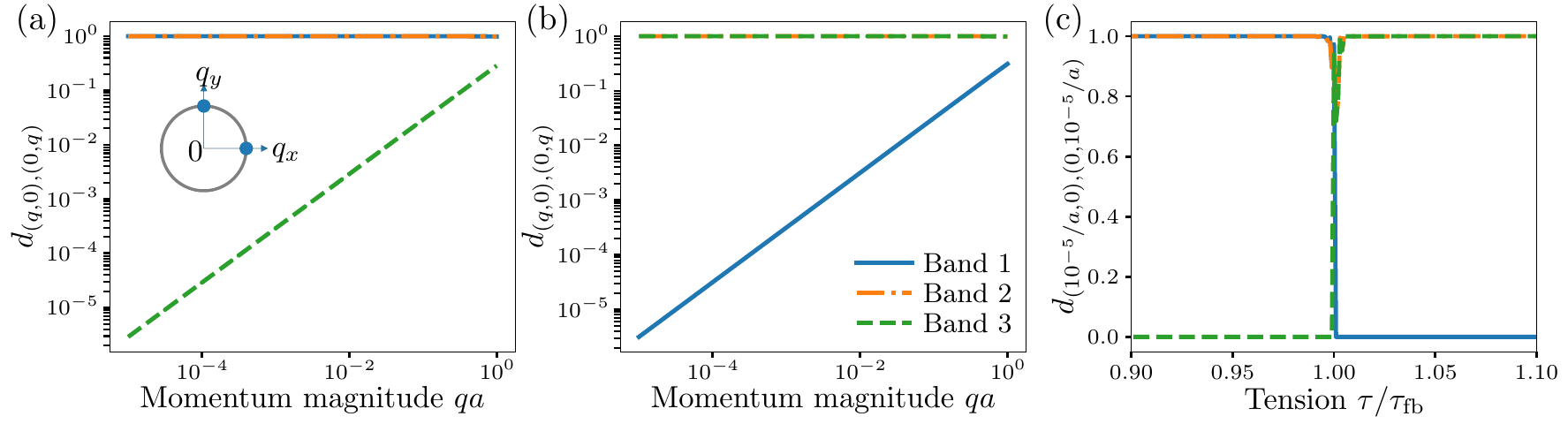}
  \caption{Hilbert-Schmidt distance computed from lowest three bands of the discrete kagome lattice model, which correspond to the vibrational states built from the fundamental modes of the three resonators in the unit cell. Bands 1--3 are ordered according to the eigenfrequency value; the legend is shared by all panels. \subfiglabel{a} Discrete model corresponding to $T < T_\text{fb}$ in \figref{Kagome-bands}, for which Band 1 is the flat band which touches Band 2 at $\bm{q}=(0,0)$. Inset shows the two crystal momenta (blue dots) at which the eigenstates are compared; the singular point is approached by setting the radius $q$ of the circle arbitrarily close to zero. \subfiglabel{b} Discrete model corresponding to $T > T_\text{fb}$ in \figref{Kagome-bands}, for which Band 3 is the flat band which touches Band 2 at $\bm{q}=(0,0)$. \subfiglabel{c} Hilbert-Schmidt distance of the three bands with $q = 10^{-5}a$ when the discrete tension value $\tau$ is varied near the value corresponding to the flat band $T = T_\text{fb}$.}
  \label{fig:hsdist}
\end{figure*}

A signature of a singular QBT is that Bloch eigenstates of different crystal momenta that are close to the QBT do not overlap, but instead differ from each other by a finite amount even as the momenta at which they are evaluated become arbitrarily close to each other. This difference is quantified by the Hilbert-Schmidt distance~\cite{Rhim_2020,Rhim-Yang_2021}
\begin{equation}
	d_{\bm{q}_1,\bm{q}_2}= \sqrt{1- \left| \left< v_{\bm{q}_1} \big| v_{\bm{q}_2} \right> \right|^2},
\end{equation}
where $v_{\bm{q}_1}$ and $v_{\bm{q}_2}$ are two eigenstates from the same band at momenta $\bm{q}_1,\bm{q}_2$ that are infinitesimally close to each other near the QBT. As the distance between the two momenta tends to zero, in the absence of a singularity the Hilbert-Schmidt distance should tend to zero as well since the inner product of two identical normalized eigenstates is unity. However, for singular flat bands the metric becomes non-zero due to the immovable discontinuity in the Bloch eigenstates at the QBT. For quantum-mechanical states, the Hilbert-Schmidt distance represents a distance between eigenstates using the metric defined by the real part of the quantum geometric tensor~\cite{Berry1989}; the imaginary part of this tensor is the Berry curvature~\cite{Xiao2010a}. In the quantum context, the Hilbert-Schmidt distance (also termed the quantum distance) is a geometric feature of Bloch states in the Hilbert space with measurable physical consequences. In particular, the maximal value of $d_{\bm{q}_1,\bm{q}_2}$ evaluated for all possible momentum pairs for the flat band has been shown to dictate the spread of Landau level energies in a magnetic field~\cite{Rhim_2020}.

To investigate whether the flat band of the kagome lattice in our phononic system also exhibits a singular QBT, we numerically evaluated the Hilbert-Schmidt distance between Bloch eigenvectors at $\bm{q}_1 = (q,0)$ and $\bm{q}_2 = (0,q)$ as $q \to 0$ for the lowest three bands of the discrete model (see inset to \subfigref{fig:hsdist}{a} for the position of the momenta). For discrete model parameters corresponding to $T < T_\text{fb}$ (\subfigref{fig:hsdist}{a}), the flat band (Band 1) and the next-highest band (Band 2) both exhibit the maximal possible value of $d_{\bm{q}_1,\bm{q}_2}=1$ as the QBT at $q=0$ is approached, showing that the QBT is singular. By contrast, the third band exhibits a Hilbert-Schmidt distance between the two states that approaches zero as $q \to 0$, which is the expected behavior in the absence of a singularity. When the tension is increased beyond the point where three flat bands arise ($T > T_\text{fb}$; \subfigref{fig:hsdist}{b}), the flat band is now third in order of frequency (Band 3) and touches Band 2 at the origin; the corresponding Hilbert-Schmidt distances again remain at one even as $q \to 0$, whereas the separated band (now the lowest band, Band 1) is non-singular. We can confirm that the singularity follows the QBT by tracking the Hilbert-Schmidt distance of the three bands for $q$ close to zero in a discrete model with $\tau$ varied near the value corresponding to $T_\text{fb}$; the maximally singular QBT is apparent on either side of the transition. Exactly at $T=T_\text{fb}$, the three bands are degenerate at all values of $\bm{q}$ and numerical evaluation generates arbitrary superpositions of Bloch eigenvectors from the three bands, so the Hilbert-Schmidt distance of each individual band cannot be computed.

\subsection{Robust boundary modes} 

\begin{figure*}[tbh]
	\noindent\includegraphics[width=\textwidth]{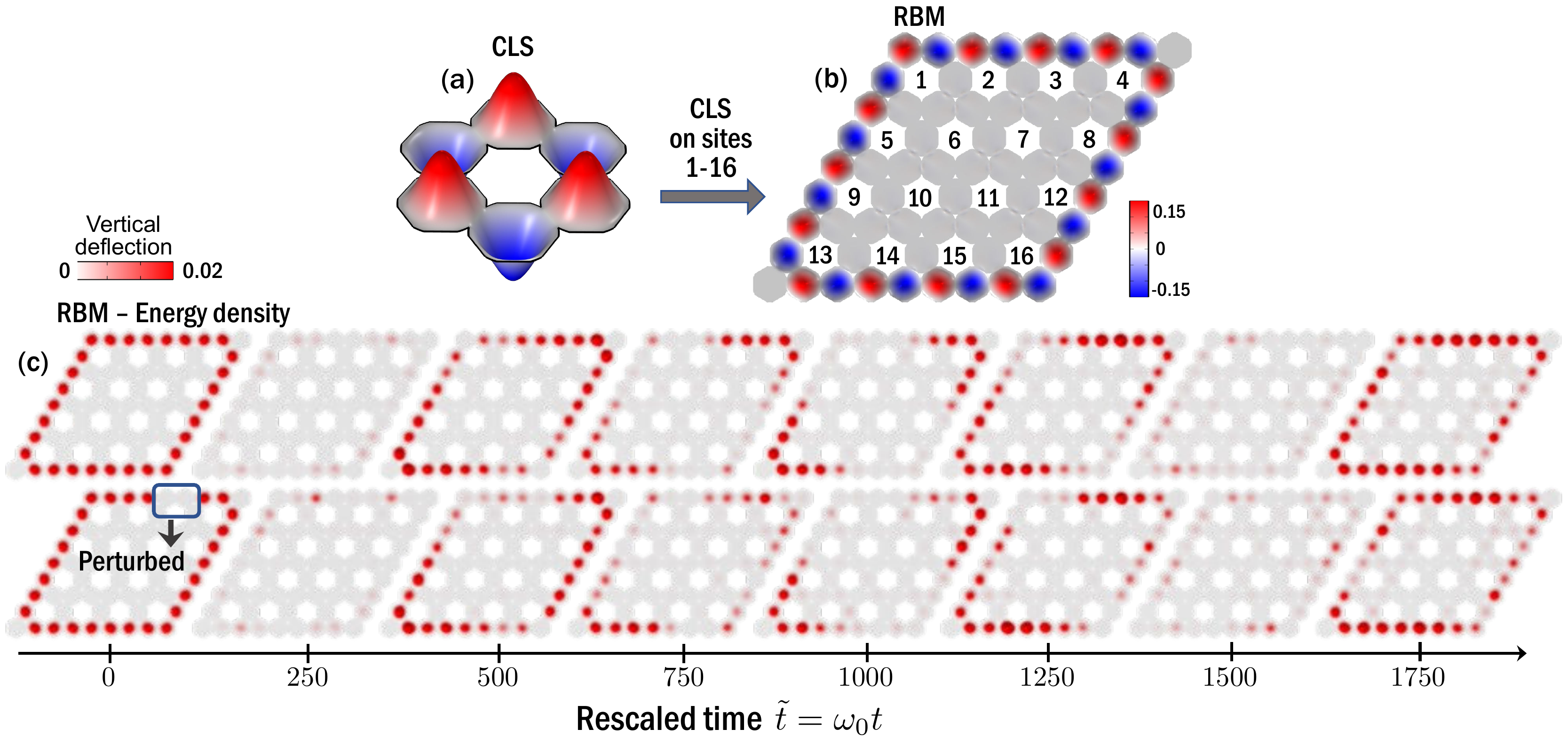}   
	\caption{   
		\label{RBMs} Robust boundary mode (RBM) created by placing the compact localized states throughout the kagome lattice. The compact localized state (a) is a flat band state and is derived from eigenmode analysis. Placing such modes around the sites labeled 1--16 in the finite section of the kagome lattice (b) creates the RBM. The destructive interference in the bulk leads to nonzero displacements only on the boundary resonators. (c) Dynamical evolution of the energy density during full-wave simulations starting fom the initial displacements in (b) (top), and with a perturbed initial condition in which the displacements on two resonators are set to zero (bottom). The tension is set to $T=1$ in the dynamical simulations. Snapshots are shown at intervals of 250 units of the rescaled time $\tilde{t}$.  
	}
\end{figure*}

The calculations of the maximal Hilbert-Schmidt distance in the vicinity of the band touching at $\bm{q}=(0,0)$ show that the discrete kagome lattice model harbors a singular flat band at tension values away from $T_\text{fb}$, and therefore supports robust boundary modes (RBMs) as the boundary manifestation of noncontractible loop states (NLSs)~\cite{Rhim_2019}. We therefore hypothesize that the continuum model also supports RBMs. To test this hypothesis, we constructed RBMs in the continuum model and investigated their dynamics according to the plate equations, \eqnref{non-dim-pde}. RBMs were constructed in the kagome lattice by placing a CLS (\subfigref{RBMs}{a}) throughout all the lattice sites of a finite section of the kagome lattice as shown in \subfigref{RBMs}{b}. In the bulk, individual CLSs interfere destructively leading to net zero displacements, with the displacements surviving only at the edges. 

Dynamical simulations reported in the top row of \subfigref{RBMs}{c} show that the RBM remains confined to the boundary over hundreds of cycles, even though the system harbors many additional states in its interior at the same frequency $\omega_0$ (the flat band frequency). This robustness arises because the RBMs are vestiges in a finite system of noncontractible loop states (NLSs)---line-like flat band states that loop around periodic directions in the absence of boundaries~\cite{Rhim_2019}. Whereas the CLSs form a spanning set for the space of eigenstates in a non-singular flat band, they do not span that space when the flat band is singular; additional NLSs are required to completely span the space of eigenstates belonging to the flat band. In a finite section of the lattice with a boundary, the NLSs manifest as closed edge states which loop around the boundary; these are the RBMs. The RBMs cannot be disconnected by adding or subtracting CLSs around the boundary; a closed loop of resonators oscillating with alternating phases is always maintained.

Motivated by explorations of RBMs in photonic systems~\cite{Ma_2020}, we verified the robustness of the RBM against being disconnected by initializing the full-wave simulation with a perturbed version of the boundary mode, with the displacement set to zero on two adjacent resonators (bottom row of \subfigref{RBMs}{c}). Despite this large perturbation of the initial state at two sites, the RBM is still recovered at long times and the perturbed sites initiated with zero displacements recover their displacements without the RBM falling apart or leaking energy to the bulk. These results suggest that singular flat bands can be exploited to generate robust edge modes in acoustic metamaterials. 

\section{Discussion}

We have theoretically and numerically investigated the vibrational properties of planar resonator lattices whose acoustic spectra are designed to harbor non-singular and singular flat bands. The acoustic band structures of the lattices can be dramatically modified by changing the in-plane mechanical tension, which enables the structures to be tuned to a state with an isolated flat band (for the triangular lattice), two degenerate flat bands (honeycomb lattice), or three degenerate flat bands, one of which remains flat at all tension values (kagome lattice). The behavior of the relevant bands is captured in a reduced discrete model of harmonic degrees of freedom on individual resonators, coupled to their neighbors via auxiliary degrees of freedom. The discrete model elucidates the mechanism underlying the lack of dispersion for certain bands at particular tension values, and establishes the singular nature of the persistent flat band in the kagome lattice. We demonstrated the presence of robust boundary modes, which derive from extended states belonging to the singular flat band that are guaranteed to exist by topology, in a finite section of a kagome lattice.  

Our study indicates that engineered plate resonator arrays could serve as a versatile tunable platform for exploring flat band physics. In contrast to photonic systems which have been used to realize exotic flat-band states~\cite{Tang2020}, mechanical systems have the advantage of being highly tunable after fabrication~\cite{Wang2020}, enabling the tuning of systems toward and away from the flat band state as our computations demonstrate. 
Experimental implementations of our proposal would require thin-plate or membrane resonator arrays with tunable in-plane tension; these have been achieved in micromechanical systems~\cite{Cha_2018, Zande_2010}. Single- or few-atomic-layer graphene membranes are suspended over voids fabricated in semiconductor substrates with the desired metamaterial geometry, and the boundaries are restricted by the adhesion of the membrane to the semiconductor at the void boundaries. Tension modulation could be achieved via electrostatic~\cite{Cha_2018,Mei_2018} or thermally-induced~\cite{Blaikie_2019} prestresses. Alternatively, resonators based on MXenes with thermally responsive rigidity~\cite{Ankit_2021} and stiffness greater than graphene~\cite{Siriwardane_2018} could serve as possible candidates for experimental realizations.

While we have highlighted robust boundary modes as a specific example of exotic flat-band physics in our system, our platform could be used or modified to investigate  other properties of flat bands as well. As we have shown, our system exhibits degenerate pairs and trios of flat bands for certain  geometries and tensions. The quantum geometry of degenerate flat bands is known to harbor unusual properties such as non-additivity of the quantum metric~\cite{Mera2022}; our system provides a classical platform in which to investigate the consequences of such nontrivial geometry of the Bloch eigenfunctions. The effect of a gauge potential, particularly a magnetic field, can be synthesized in the vibrational spectrum of mechanical systems via spatial modulation of the resonator geometry or tension~\cite{Yang2017,Abbaszadeh2016,Brendel2017,Wen2019,Yan2021}; using such techniques, the interplay of magnetic fields and singular flat bands~\cite{Rhim_2020} could be investigated in mechanical resonator networks.

Beyond questions of fundamental interest, the exotic vibrational states enabled by flat bands could serve as the basis of acoustic metamaterials with useful properties. The ability to tune the dispersion towards and away from a flat band using tension modulation could be used to manipulate or arrest the motion of sound pulses~\cite{Karki_2021}. Periodic modulation of the background tension near the flat band value could be used to parametrically amplify vibrations in flat-band states~\cite{Rugar1991}; since all states in the flat band would resonate at the same parametric modulation frequency, multispectral signals could be amplified without loss of fidelity. The doubly-degenerate flat bands in the honeycomb lattice can be used to engineer vibrational states with spin-like degrees of freedom that exhibit non-Abelian responses, which have been proposed as a platform for mechanical computation~\cite{Fruchart2020}. More generally, our work suggests that the collective vibrations of coupled resonators can be used to implement mechanical analogues of two-dimensional tight-binding models~\cite{Matlack2018a}, with the possibility of controlling the strength and sign of the coupling terms~\cite{Keil2016} through tension modulation after fabrication.

\section{Acknowledgments}

We acknowledge support from the College of Arts and
Sciences at the University of Oregon via startup funds to JP. Work was partially
supported by the National Science Foundation under Grant~No.~CMMI-2128671.

\appendix

\section{Finite-element analysis of continuum model \label{appendixFiniteElementAnalysis}}

Finite-element analyses were done in the commercially available package COMSOL Multiphysics.
The \emph{general form pde} module was used to define an eigenvalue problem based on a
fourth-order partial differential equation describing thin plate elasticity,
\begin{equation} 
	\nabla \cdot  \bigg[ \left(  u_{xxx} + 2 u_{xyy} - T u_x \right)  \hat{x} + \left(  u_{yyy}  - T u_y \right) \hat{y} \bigg]  =\lambda u,
	\label{FEM}
\end{equation}
where subscripts denote partial derivatives of $u$ with respect to those coordinates.
The Dirichlet boundary condition $u=0$ and Neumann boundary conditions $u_x=u_y=0$ satisfy the clamped 
boundary condition. Simplifying the equation \eqref{FEM} gives
$\nabla^4 u-T\nabla^2 u = \lambda u$, which is the desired eigenvalue problem. 
The simulation methods were tested by comparing numerically-derived eigensolutions 
for the square Laplacian plate (setting $D\rightarrow0$) and circular clamped biharmonic plate (setting
$T\rightarrow0$) to known analytical results. 

The continuum thin-plate resonator model has infinitely many bands, of which a
subset are obtained numerically. In this study, we focused solely on the bands 
corresponding to the fundamental modes, which are built
primarily from the lowest-frequency modes of individual
resonators. This is apparent from the mode shape of the Bloch eigenfunction 
in the continuum model which mirrors the mode shape of the
single-resonator fundamental mode.

The full-wave dynamical simulations were performed by adding the second-order time derivative term to the 
partial differential equation,
\begin{equation}
\frac{\partial^2 u}{\partial t^2}+\nabla^4 u-T\nabla^2 u=0. 
\end{equation}

\textit{Mesh details for FEA:}
A custom mesh in COMSOL was used for the finite element analysis. Element size parameters for the mesh are as follows: maximum element size = $0.08$, minimum element size = $1.5\times10^{-4}$, maximum element growth rate = 1.2, curvature factor = 0.25, and resolution of
narrow regions = 1.

A time-step of 0.01 was used in the time-dependent solver called the generalized alpha. 


\input{2DFlatBands.bbl}

\end{document}

%% file: 2DFlatBands.bbl
%

%% file: 2DFlatBands.bbl
\begin{thebibliography}{61}%
\makeatletter
\providecommand \@ifxundefined [1]{%
 \@ifx{#1\undefined}
}%
\providecommand \@ifnum [1]{%
 \ifnum #1\expandafter \@firstoftwo
 \else \expandafter \@secondoftwo
 \fi
}%
\providecommand \@ifx [1]{%
 \ifx #1\expandafter \@firstoftwo
 \else \expandafter \@secondoftwo
 \fi
}%
\providecommand \natexlab [1]{#1}%
\providecommand \enquote  [1]{``#1''}%
\providecommand \bibnamefont  [1]{#1}%
\providecommand \bibfnamefont [1]{#1}%
\providecommand \citenamefont [1]{#1}%
\providecommand \href@noop [0]{\@secondoftwo}%
\providecommand \href [0]{\begingroup \@sanitize@url \@href}%
\providecommand \@href[1]{\@@startlink{#1}\@@href}%
\providecommand \@@href[1]{\endgroup#1\@@endlink}%
\providecommand \@sanitize@url [0]{\catcode `\\12\catcode `\$12\catcode
  `\&12\catcode `\#12\catcode `\^12\catcode `\_12\catcode `\%12\relax}%
\providecommand \@@startlink[1]{}%
\providecommand \@@endlink[0]{}%
\providecommand \url  [0]{\begingroup\@sanitize@url \@url }%
\providecommand \@url [1]{\endgroup\@href {#1}{\urlprefix }}%
\providecommand \urlprefix  [0]{URL }%
\providecommand \Eprint [0]{\href }%
\providecommand \doibase [0]{https://doi.org/}%
\providecommand \selectlanguage [0]{\@gobble}%
\providecommand \bibinfo  [0]{\@secondoftwo}%
\providecommand \bibfield  [0]{\@secondoftwo}%
\providecommand \translation [1]{[#1]}%
\providecommand \BibitemOpen [0]{}%
\providecommand \bibitemStop [0]{}%
\providecommand \bibitemNoStop [0]{.\EOS\space}%
\providecommand \EOS [0]{\spacefactor3000\relax}%
\providecommand \BibitemShut  [1]{\csname bibitem#1\endcsname}%
\let\auto@bib@innerbib\@empty
\bibitem [{\citenamefont {Sutherland}(1986)}]{Sutherland_1986}%
  \BibitemOpen
  \bibfield  {author} {\bibinfo {author} {\bibfnamefont {B.}~\bibnamefont
  {Sutherland}},\ }\bibfield  {title} {\bibinfo {title} {Localization of
  electronic wave functions due to local topology},\ }\href
  {https://doi.org/10.1103/PhysRevB.34.5208} {\bibfield  {journal} {\bibinfo
  {journal} {Phys. Rev. B}\ }\textbf {\bibinfo {volume} {34}},\ \bibinfo
  {pages} {5208} (\bibinfo {year} {1986})}\BibitemShut {NoStop}%
\bibitem [{\citenamefont {Mielke}(1991)}]{Mielke_1991}%
  \BibitemOpen
  \bibfield  {author} {\bibinfo {author} {\bibfnamefont {A.}~\bibnamefont
  {Mielke}},\ }\bibfield  {title} {\bibinfo {title} {Ferromagnetic ground
  states for the hubbard model on line graphs},\ }\href
  {https://doi.org/10.1088/0305-4470/24/2/005} {\bibfield  {journal} {\bibinfo
  {journal} {Journal of Physics A: Mathematical and General}\ }\textbf
  {\bibinfo {volume} {24}},\ \bibinfo {pages} {L73} (\bibinfo {year}
  {1991})}\BibitemShut {NoStop}%
\bibitem [{\citenamefont {Tasaki}(1994)}]{Tasaki_1994}%
  \BibitemOpen
  \bibfield  {author} {\bibinfo {author} {\bibfnamefont {H.}~\bibnamefont
  {Tasaki}},\ }\bibfield  {title} {\bibinfo {title} {Stability of
  ferromagnetism in the hubbard model},\ }\href
  {https://doi.org/10.1103/PhysRevLett.73.1158} {\bibfield  {journal} {\bibinfo
   {journal} {Phys. Rev. Lett.}\ }\textbf {\bibinfo {volume} {73}},\ \bibinfo
  {pages} {1158} (\bibinfo {year} {1994})}\BibitemShut {NoStop}%
\bibitem [{\citenamefont {Tamura}\ \emph {et~al.}(2002)\citenamefont {Tamura},
  \citenamefont {Shiraishi}, \citenamefont {Kimura},\ and\ \citenamefont
  {Takayanagi}}]{Tamura_2002}%
  \BibitemOpen
  \bibfield  {author} {\bibinfo {author} {\bibfnamefont {H.}~\bibnamefont
  {Tamura}}, \bibinfo {author} {\bibfnamefont {K.}~\bibnamefont {Shiraishi}},
  \bibinfo {author} {\bibfnamefont {T.}~\bibnamefont {Kimura}},\ and\ \bibinfo
  {author} {\bibfnamefont {H.}~\bibnamefont {Takayanagi}},\ }\bibfield  {title}
  {\bibinfo {title} {Flat-band ferromagnetism in quantum dot superlattices},\
  }\href {https://doi.org/10.1103/PhysRevB.65.085324} {\bibfield  {journal}
  {\bibinfo  {journal} {Phys. Rev. B}\ }\textbf {\bibinfo {volume} {65}},\
  \bibinfo {pages} {085324} (\bibinfo {year} {2002})}\BibitemShut {NoStop}%
\bibitem [{\citenamefont {Maksymenko}\ \emph {et~al.}(2012)\citenamefont
  {Maksymenko}, \citenamefont {Honecker}, \citenamefont {Moessner},
  \citenamefont {Richter},\ and\ \citenamefont {Derzhko}}]{Maksymenko_2012}%
  \BibitemOpen
  \bibfield  {author} {\bibinfo {author} {\bibfnamefont {M.}~\bibnamefont
  {Maksymenko}}, \bibinfo {author} {\bibfnamefont {A.}~\bibnamefont
  {Honecker}}, \bibinfo {author} {\bibfnamefont {R.}~\bibnamefont {Moessner}},
  \bibinfo {author} {\bibfnamefont {J.}~\bibnamefont {Richter}},\ and\ \bibinfo
  {author} {\bibfnamefont {O.}~\bibnamefont {Derzhko}},\ }\bibfield  {title}
  {\bibinfo {title} {Flat-band ferromagnetism as a pauli-correlated percolation
  problem},\ }\href {https://doi.org/10.1103/PhysRevLett.109.096404} {\bibfield
   {journal} {\bibinfo  {journal} {Phys. Rev. Lett.}\ }\textbf {\bibinfo
  {volume} {109}},\ \bibinfo {pages} {096404} (\bibinfo {year}
  {2012})}\BibitemShut {NoStop}%
\bibitem [{\citenamefont {Hase}\ \emph {et~al.}(2018)\citenamefont {Hase},
  \citenamefont {Yanagisawa}, \citenamefont {Aiura},\ and\ \citenamefont
  {Kawashima}}]{Hase_2018}%
  \BibitemOpen
  \bibfield  {author} {\bibinfo {author} {\bibfnamefont {I.}~\bibnamefont
  {Hase}}, \bibinfo {author} {\bibfnamefont {T.}~\bibnamefont {Yanagisawa}},
  \bibinfo {author} {\bibfnamefont {Y.}~\bibnamefont {Aiura}},\ and\ \bibinfo
  {author} {\bibfnamefont {K.}~\bibnamefont {Kawashima}},\ }\bibfield  {title}
  {\bibinfo {title} {Possibility of flat-band ferromagnetism in hole-doped
  pyrochlore oxides ${\mathrm{sn}}_{2}{\mathrm{nb}}_{2}{\mathrm{o}}_{7}$ and
  ${\mathrm{sn}}_{2}{\mathrm{ta}}_{2}{\mathrm{o}}_{7}$},\ }\href
  {https://doi.org/10.1103/PhysRevLett.120.196401} {\bibfield  {journal}
  {\bibinfo  {journal} {Phys. Rev. Lett.}\ }\textbf {\bibinfo {volume} {120}},\
  \bibinfo {pages} {196401} (\bibinfo {year} {2018})}\BibitemShut {NoStop}%
\bibitem [{\citenamefont {Misumi}\ and\ \citenamefont
  {Aoki}(2017)}]{Misumi_2017}%
  \BibitemOpen
  \bibfield  {author} {\bibinfo {author} {\bibfnamefont {T.}~\bibnamefont
  {Misumi}}\ and\ \bibinfo {author} {\bibfnamefont {H.}~\bibnamefont {Aoki}},\
  }\bibfield  {title} {\bibinfo {title} {New class of flat-band models on
  tetragonal and hexagonal lattices: Gapped versus crossing flat bands},\
  }\href {https://doi.org/10.1103/PhysRevB.96.155137} {\bibfield  {journal}
  {\bibinfo  {journal} {Phys. Rev. B}\ }\textbf {\bibinfo {volume} {96}},\
  \bibinfo {pages} {155137} (\bibinfo {year} {2017})}\BibitemShut {NoStop}%
\bibitem [{\citenamefont {Zyuzin}\ and\ \citenamefont
  {Zyuzin}(2018)}]{Zyuzin_2018}%
  \BibitemOpen
  \bibfield  {author} {\bibinfo {author} {\bibfnamefont {A.~A.}\ \bibnamefont
  {Zyuzin}}\ and\ \bibinfo {author} {\bibfnamefont {A.~Y.}\ \bibnamefont
  {Zyuzin}},\ }\bibfield  {title} {\bibinfo {title} {Flat band in
  disorder-driven non-hermitian weyl semimetals},\ }\href
  {https://doi.org/10.1103/PhysRevB.97.041203} {\bibfield  {journal} {\bibinfo
  {journal} {Phys. Rev. B}\ }\textbf {\bibinfo {volume} {97}},\ \bibinfo
  {pages} {041203} (\bibinfo {year} {2018})}\BibitemShut {NoStop}%
\bibitem [{\citenamefont {Kumar}\ \emph {et~al.}(2021)\citenamefont {Kumar},
  \citenamefont {Chen},\ and\ \citenamefont {Lado}}]{Kumar_2021}%
  \BibitemOpen
  \bibfield  {author} {\bibinfo {author} {\bibfnamefont {P.}~\bibnamefont
  {Kumar}}, \bibinfo {author} {\bibfnamefont {G.}~\bibnamefont {Chen}},\ and\
  \bibinfo {author} {\bibfnamefont {J.~L.}\ \bibnamefont {Lado}},\ }\bibfield
  {title} {\bibinfo {title} {Kondo lattice mediated interactions in flat-band
  systems},\ }\href {https://doi.org/10.1103/PhysRevResearch.3.043113}
  {\bibfield  {journal} {\bibinfo  {journal} {Phys. Rev. Research}\ }\textbf
  {\bibinfo {volume} {3}},\ \bibinfo {pages} {043113} (\bibinfo {year}
  {2021})}\BibitemShut {NoStop}%
\bibitem [{\citenamefont {Morfonios}\ \emph {et~al.}(2021)\citenamefont
  {Morfonios}, \citenamefont {R\"ontgen}, \citenamefont {Pyzh},\ and\
  \citenamefont {Schmelcher}}]{Morfonios_2021}%
  \BibitemOpen
  \bibfield  {author} {\bibinfo {author} {\bibfnamefont {C.~V.}\ \bibnamefont
  {Morfonios}}, \bibinfo {author} {\bibfnamefont {M.}~\bibnamefont
  {R\"ontgen}}, \bibinfo {author} {\bibfnamefont {M.}~\bibnamefont {Pyzh}},\
  and\ \bibinfo {author} {\bibfnamefont {P.}~\bibnamefont {Schmelcher}},\
  }\bibfield  {title} {\bibinfo {title} {Flat bands by latent symmetry},\
  }\href {https://doi.org/10.1103/PhysRevB.104.035105} {\bibfield  {journal}
  {\bibinfo  {journal} {Phys. Rev. B}\ }\textbf {\bibinfo {volume} {104}},\
  \bibinfo {pages} {035105} (\bibinfo {year} {2021})}\BibitemShut {NoStop}%
\bibitem [{\citenamefont {Shen}\ \emph {et~al.}(2010)\citenamefont {Shen},
  \citenamefont {Shao}, \citenamefont {Wang},\ and\ \citenamefont
  {Xing}}]{Shen_2010}%
  \BibitemOpen
  \bibfield  {author} {\bibinfo {author} {\bibfnamefont {R.}~\bibnamefont
  {Shen}}, \bibinfo {author} {\bibfnamefont {L.~B.}\ \bibnamefont {Shao}},
  \bibinfo {author} {\bibfnamefont {B.}~\bibnamefont {Wang}},\ and\ \bibinfo
  {author} {\bibfnamefont {D.~Y.}\ \bibnamefont {Xing}},\ }\bibfield  {title}
  {\bibinfo {title} {Single dirac cone with a flat band touching on
  line-centered-square optical lattices},\ }\href
  {https://doi.org/10.1103/PhysRevB.81.041410} {\bibfield  {journal} {\bibinfo
  {journal} {Phys. Rev. B}\ }\textbf {\bibinfo {volume} {81}},\ \bibinfo
  {pages} {041410} (\bibinfo {year} {2010})}\BibitemShut {NoStop}%
\bibitem [{\citenamefont {Apaja}\ \emph {et~al.}(2010)\citenamefont {Apaja},
  \citenamefont {Hyrk\"as},\ and\ \citenamefont {Manninen}}]{Apaja_2010}%
  \BibitemOpen
  \bibfield  {author} {\bibinfo {author} {\bibfnamefont {V.}~\bibnamefont
  {Apaja}}, \bibinfo {author} {\bibfnamefont {M.}~\bibnamefont {Hyrk\"as}},\
  and\ \bibinfo {author} {\bibfnamefont {M.}~\bibnamefont {Manninen}},\
  }\bibfield  {title} {\bibinfo {title} {Flat bands, dirac cones, and atom
  dynamics in an optical lattice},\ }\href
  {https://doi.org/10.1103/PhysRevA.82.041402} {\bibfield  {journal} {\bibinfo
  {journal} {Phys. Rev. A}\ }\textbf {\bibinfo {volume} {82}},\ \bibinfo
  {pages} {041402} (\bibinfo {year} {2010})}\BibitemShut {NoStop}%
\bibitem [{\citenamefont {Mukherjee}\ \emph {et~al.}(2015)\citenamefont
  {Mukherjee}, \citenamefont {Spracklen}, \citenamefont {Choudhury},
  \citenamefont {Goldman}, \citenamefont {\"Ohberg}, \citenamefont
  {Andersson},\ and\ \citenamefont {Thomson}}]{Mukherjee_2015}%
  \BibitemOpen
  \bibfield  {author} {\bibinfo {author} {\bibfnamefont {S.}~\bibnamefont
  {Mukherjee}}, \bibinfo {author} {\bibfnamefont {A.}~\bibnamefont
  {Spracklen}}, \bibinfo {author} {\bibfnamefont {D.}~\bibnamefont
  {Choudhury}}, \bibinfo {author} {\bibfnamefont {N.}~\bibnamefont {Goldman}},
  \bibinfo {author} {\bibfnamefont {P.}~\bibnamefont {\"Ohberg}}, \bibinfo
  {author} {\bibfnamefont {E.}~\bibnamefont {Andersson}},\ and\ \bibinfo
  {author} {\bibfnamefont {R.~R.}\ \bibnamefont {Thomson}},\ }\bibfield
  {title} {\bibinfo {title} {Observation of a localized flat-band state in a
  photonic lieb lattice},\ }\href
  {https://doi.org/10.1103/PhysRevLett.114.245504} {\bibfield  {journal}
  {\bibinfo  {journal} {Phys. Rev. Lett.}\ }\textbf {\bibinfo {volume} {114}},\
  \bibinfo {pages} {245504} (\bibinfo {year} {2015})}\BibitemShut {NoStop}%
\bibitem [{\citenamefont {Travkin}\ \emph {et~al.}(2017)\citenamefont
  {Travkin}, \citenamefont {Diebel},\ and\ \citenamefont
  {Denz}}]{Travkin_2017}%
  \BibitemOpen
  \bibfield  {author} {\bibinfo {author} {\bibfnamefont {E.}~\bibnamefont
  {Travkin}}, \bibinfo {author} {\bibfnamefont {F.}~\bibnamefont {Diebel}},\
  and\ \bibinfo {author} {\bibfnamefont {C.}~\bibnamefont {Denz}},\ }\bibfield
  {title} {\bibinfo {title} {Compact flat band states in optically induced
  flatland photonic lattices},\ }\href {https://doi.org/10.1063/1.4990998}
  {\bibfield  {journal} {\bibinfo  {journal} {Applied Physics Letters}\
  }\textbf {\bibinfo {volume} {111}},\ \bibinfo {pages} {011104} (\bibinfo
  {year} {2017})}\BibitemShut {NoStop}%
\bibitem [{\citenamefont {Ramezani}(2017)}]{Hamidreza_2017}%
  \BibitemOpen
  \bibfield  {author} {\bibinfo {author} {\bibfnamefont {H.}~\bibnamefont
  {Ramezani}},\ }\bibfield  {title} {\bibinfo {title} {Non-hermiticity-induced
  flat band},\ }\href {https://doi.org/10.1103/PhysRevA.96.011802} {\bibfield
  {journal} {\bibinfo  {journal} {Phys. Rev. A}\ }\textbf {\bibinfo {volume}
  {96}},\ \bibinfo {pages} {011802} (\bibinfo {year} {2017})}\BibitemShut
  {NoStop}%
\bibitem [{\citenamefont {Ge}(2018)}]{Ge_2018}%
  \BibitemOpen
  \bibfield  {author} {\bibinfo {author} {\bibfnamefont {L.}~\bibnamefont
  {Ge}},\ }\bibfield  {title} {\bibinfo {title} {Non-hermitian lattices with a
  flat band and polynomial power increase [invited]},\ }\href
  {https://doi.org/10.1364/PRJ.6.000A10} {\bibfield  {journal} {\bibinfo
  {journal} {Photon. Res.}\ }\textbf {\bibinfo {volume} {6}},\ \bibinfo {pages}
  {A10} (\bibinfo {year} {2018})}\BibitemShut {NoStop}%
\bibitem [{\citenamefont {Longhi}(2019)}]{Longhi_2019}%
  \BibitemOpen
  \bibfield  {author} {\bibinfo {author} {\bibfnamefont {S.}~\bibnamefont
  {Longhi}},\ }\bibfield  {title} {\bibinfo {title} {Photonic flat-band
  laser},\ }\href {https://doi.org/10.1364/OL.44.000287} {\bibfield  {journal}
  {\bibinfo  {journal} {Opt. Lett.}\ }\textbf {\bibinfo {volume} {44}},\
  \bibinfo {pages} {287} (\bibinfo {year} {2019})}\BibitemShut {NoStop}%
\bibitem [{\citenamefont {Zheng}\ \emph {et~al.}(2014)\citenamefont {Zheng},
  \citenamefont {Wu}, \citenamefont {Ni}, \citenamefont {Chen}, \citenamefont
  {Lu},\ and\ \citenamefont {Chen}}]{Zheng_2014}%
  \BibitemOpen
  \bibfield  {author} {\bibinfo {author} {\bibfnamefont {L.-Y.}\ \bibnamefont
  {Zheng}}, \bibinfo {author} {\bibfnamefont {Y.}~\bibnamefont {Wu}}, \bibinfo
  {author} {\bibfnamefont {X.}~\bibnamefont {Ni}}, \bibinfo {author}
  {\bibfnamefont {Z.-G.}\ \bibnamefont {Chen}}, \bibinfo {author}
  {\bibfnamefont {M.-H.}\ \bibnamefont {Lu}},\ and\ \bibinfo {author}
  {\bibfnamefont {Y.-F.}\ \bibnamefont {Chen}},\ }\bibfield  {title} {\bibinfo
  {title} {Acoustic cloaking by a near-zero-index phononic crystal},\ }\href
  {https://doi.org/10.1063/1.4873354} {\bibfield  {journal} {\bibinfo
  {journal} {Applied Physics Letters}\ }\textbf {\bibinfo {volume} {104}},\
  \bibinfo {pages} {161904} (\bibinfo {year} {2014})}\BibitemShut {NoStop}%
\bibitem [{\citenamefont {Dubois}\ \emph {et~al.}(2019)\citenamefont {Dubois},
  \citenamefont {Perchoux}, \citenamefont {Vanel}, \citenamefont {Tronche},
  \citenamefont {Achaoui}, \citenamefont {Dupont}, \citenamefont {Bertling},
  \citenamefont {Raki\ifmmode~\acute{c}\else \'{c}\fi{}}, \citenamefont
  {Antonakakis}, \citenamefont {Enoch}, \citenamefont {Abdeddaim},
  \citenamefont {Craster},\ and\ \citenamefont {Guenneau}}]{Dubois-2019}%
  \BibitemOpen
  \bibfield  {author} {\bibinfo {author} {\bibfnamefont {M.}~\bibnamefont
  {Dubois}}, \bibinfo {author} {\bibfnamefont {J.}~\bibnamefont {Perchoux}},
  \bibinfo {author} {\bibfnamefont {A.~L.}\ \bibnamefont {Vanel}}, \bibinfo
  {author} {\bibfnamefont {C.}~\bibnamefont {Tronche}}, \bibinfo {author}
  {\bibfnamefont {Y.}~\bibnamefont {Achaoui}}, \bibinfo {author} {\bibfnamefont
  {G.}~\bibnamefont {Dupont}}, \bibinfo {author} {\bibfnamefont
  {K.}~\bibnamefont {Bertling}}, \bibinfo {author} {\bibfnamefont {A.~D.}\
  \bibnamefont {Raki\ifmmode~\acute{c}\else \'{c}\fi{}}}, \bibinfo {author}
  {\bibfnamefont {T.}~\bibnamefont {Antonakakis}}, \bibinfo {author}
  {\bibfnamefont {S.}~\bibnamefont {Enoch}}, \bibinfo {author} {\bibfnamefont
  {R.}~\bibnamefont {Abdeddaim}}, \bibinfo {author} {\bibfnamefont {R.~V.}\
  \bibnamefont {Craster}},\ and\ \bibinfo {author} {\bibfnamefont
  {S.}~\bibnamefont {Guenneau}},\ }\bibfield  {title} {\bibinfo {title}
  {Acoustic flat lensing using an indefinite medium},\ }\href
  {https://doi.org/10.1103/PhysRevB.99.100301} {\bibfield  {journal} {\bibinfo
  {journal} {Phys. Rev. B}\ }\textbf {\bibinfo {volume} {99}},\ \bibinfo
  {pages} {100301} (\bibinfo {year} {2019})}\BibitemShut {NoStop}%
\bibitem [{\citenamefont {Wu}\ and\ \citenamefont {Mei}(2016)}]{Wu_2016}%
  \BibitemOpen
  \bibfield  {author} {\bibinfo {author} {\bibfnamefont {S.}~\bibnamefont
  {Wu}}\ and\ \bibinfo {author} {\bibfnamefont {J.}~\bibnamefont {Mei}},\
  }\bibfield  {title} {\bibinfo {title} {Flat band degeneracy and near-zero
  refractive index materials in acoustic crystals},\ }\href
  {https://doi.org/10.1063/1.4939847} {\bibfield  {journal} {\bibinfo
  {journal} {AIP Advances}\ }\textbf {\bibinfo {volume} {6}},\ \bibinfo {pages}
  {015204} (\bibinfo {year} {2016})}\BibitemShut {NoStop}%
\bibitem [{\citenamefont {Shen}\ \emph {et~al.}(2022)\citenamefont {Shen},
  \citenamefont {Peng}, \citenamefont {Cao}, \citenamefont {Li},\ and\
  \citenamefont {Zhu}}]{Shen-2022}%
  \BibitemOpen
  \bibfield  {author} {\bibinfo {author} {\bibfnamefont {Y.-X.}\ \bibnamefont
  {Shen}}, \bibinfo {author} {\bibfnamefont {Y.-G.}\ \bibnamefont {Peng}},
  \bibinfo {author} {\bibfnamefont {P.-C.}\ \bibnamefont {Cao}}, \bibinfo
  {author} {\bibfnamefont {J.}~\bibnamefont {Li}},\ and\ \bibinfo {author}
  {\bibfnamefont {X.-F.}\ \bibnamefont {Zhu}},\ }\bibfield  {title} {\bibinfo
  {title} {Observing localization and delocalization of the flat-band states in
  an acoustic cubic lattice},\ }\href
  {https://doi.org/10.1103/PhysRevB.105.104102} {\bibfield  {journal} {\bibinfo
   {journal} {Phys. Rev. B}\ }\textbf {\bibinfo {volume} {105}},\ \bibinfo
  {pages} {104102} (\bibinfo {year} {2022})}\BibitemShut {NoStop}%
\bibitem [{\citenamefont {Xia}\ \emph {et~al.}(2016)\citenamefont {Xia},
  \citenamefont {Hu}, \citenamefont {Song}, \citenamefont {Zong}, \citenamefont
  {Tang},\ and\ \citenamefont {Chen}}]{Xia-2016}%
  \BibitemOpen
  \bibfield  {author} {\bibinfo {author} {\bibfnamefont {S.}~\bibnamefont
  {Xia}}, \bibinfo {author} {\bibfnamefont {Y.}~\bibnamefont {Hu}}, \bibinfo
  {author} {\bibfnamefont {D.}~\bibnamefont {Song}}, \bibinfo {author}
  {\bibfnamefont {Y.}~\bibnamefont {Zong}}, \bibinfo {author} {\bibfnamefont
  {L.}~\bibnamefont {Tang}},\ and\ \bibinfo {author} {\bibfnamefont
  {Z.}~\bibnamefont {Chen}},\ }\bibfield  {title} {\bibinfo {title}
  {Demonstration of flat-band image transmission in optically induced lieb
  photonic lattices},\ }\href {https://doi.org/10.1364/OL.41.001435} {\bibfield
   {journal} {\bibinfo  {journal} {Opt. Lett.}\ }\textbf {\bibinfo {volume}
  {41}},\ \bibinfo {pages} {1435} (\bibinfo {year} {2016})}\BibitemShut
  {NoStop}%
\bibitem [{\citenamefont {Real}\ \emph {et~al.}(2017)\citenamefont {Real},
  \citenamefont {Cantillano}, \citenamefont {L{\'o}pez-Gonz{\'a}lez},
  \citenamefont {Szameit}, \citenamefont {Aono}, \citenamefont {Naruse},
  \citenamefont {Kim}, \citenamefont {Wang},\ and\ \citenamefont
  {Vicencio}}]{Bastian-2017}%
  \BibitemOpen
  \bibfield  {author} {\bibinfo {author} {\bibfnamefont {B.}~\bibnamefont
  {Real}}, \bibinfo {author} {\bibfnamefont {C.}~\bibnamefont {Cantillano}},
  \bibinfo {author} {\bibfnamefont {D.}~\bibnamefont {L{\'o}pez-Gonz{\'a}lez}},
  \bibinfo {author} {\bibfnamefont {A.}~\bibnamefont {Szameit}}, \bibinfo
  {author} {\bibfnamefont {M.}~\bibnamefont {Aono}}, \bibinfo {author}
  {\bibfnamefont {M.}~\bibnamefont {Naruse}}, \bibinfo {author} {\bibfnamefont
  {S.-J.}\ \bibnamefont {Kim}}, \bibinfo {author} {\bibfnamefont
  {K.}~\bibnamefont {Wang}},\ and\ \bibinfo {author} {\bibfnamefont {R.~A.}\
  \bibnamefont {Vicencio}},\ }\bibfield  {title} {\bibinfo {title} {Flat-band
  light dynamics in stub photonic lattices},\ }\href
  {https://doi.org/10.1038/s41598-017-15441-2} {\bibfield  {journal} {\bibinfo
  {journal} {Scientific Reports}\ }\textbf {\bibinfo {volume} {7}},\ \bibinfo
  {pages} {15085} (\bibinfo {year} {2017})}\BibitemShut {NoStop}%
\bibitem [{\citenamefont {Li}\ \emph {et~al.}(2008)\citenamefont {Li},
  \citenamefont {White}, \citenamefont {O'Faolain}, \citenamefont
  {Gomez-Iglesias},\ and\ \citenamefont {Krauss}}]{Li-2008}%
  \BibitemOpen
  \bibfield  {author} {\bibinfo {author} {\bibfnamefont {J.}~\bibnamefont
  {Li}}, \bibinfo {author} {\bibfnamefont {T.~P.}\ \bibnamefont {White}},
  \bibinfo {author} {\bibfnamefont {L.}~\bibnamefont {O'Faolain}}, \bibinfo
  {author} {\bibfnamefont {A.}~\bibnamefont {Gomez-Iglesias}},\ and\ \bibinfo
  {author} {\bibfnamefont {T.~F.}\ \bibnamefont {Krauss}},\ }\bibfield  {title}
  {\bibinfo {title} {Systematic design of flat band slow light in photonic
  crystal waveguides},\ }\href {https://doi.org/10.1364/OE.16.006227}
  {\bibfield  {journal} {\bibinfo  {journal} {Opt. Express}\ }\textbf {\bibinfo
  {volume} {16}},\ \bibinfo {pages} {6227} (\bibinfo {year}
  {2008})}\BibitemShut {NoStop}%
\bibitem [{\citenamefont {Kim}\ and\ \citenamefont {Kim}(2022)}]{Kim-2022}%
  \BibitemOpen
  \bibfield  {author} {\bibinfo {author} {\bibfnamefont {K.}~\bibnamefont
  {Kim}}\ and\ \bibinfo {author} {\bibfnamefont {S.}~\bibnamefont {Kim}},\
  }\bibfield  {title} {\bibinfo {title} {Mode conversion and resonant
  absorption in inhomogeneous materials with flat bands},\ }\href
  {https://doi.org/10.1103/PhysRevB.105.045136} {\bibfield  {journal} {\bibinfo
   {journal} {Phys. Rev. B}\ }\textbf {\bibinfo {volume} {105}},\ \bibinfo
  {pages} {045136} (\bibinfo {year} {2022})}\BibitemShut {NoStop}%
\bibitem [{\citenamefont {Rhim}\ and\ \citenamefont {Yang}(2019)}]{Rhim_2019}%
  \BibitemOpen
  \bibfield  {author} {\bibinfo {author} {\bibfnamefont {J.-W.}\ \bibnamefont
  {Rhim}}\ and\ \bibinfo {author} {\bibfnamefont {B.-J.}\ \bibnamefont
  {Yang}},\ }\bibfield  {title} {\bibinfo {title} {Classification of flat bands
  according to the band-crossing singularity of bloch wave functions},\ }\href
  {https://doi.org/10.1103/PhysRevB.99.045107} {\bibfield  {journal} {\bibinfo
  {journal} {Phys. Rev. B}\ }\textbf {\bibinfo {volume} {99}},\ \bibinfo
  {pages} {045107} (\bibinfo {year} {2019})}\BibitemShut {NoStop}%
\bibitem [{\citenamefont {Rhim}\ and\ \citenamefont
  {Yang}(2021)}]{Rhim-Yang_2021}%
  \BibitemOpen
  \bibfield  {author} {\bibinfo {author} {\bibfnamefont {J.-W.}\ \bibnamefont
  {Rhim}}\ and\ \bibinfo {author} {\bibfnamefont {B.-J.}\ \bibnamefont
  {Yang}},\ }\bibfield  {title} {\bibinfo {title} {Singular flat bands},\
  }\href {https://doi.org/10.1080/23746149.2021.1901606} {\bibfield  {journal}
  {\bibinfo  {journal} {Advances in Physics: X}\ }\textbf {\bibinfo {volume}
  {6}},\ \bibinfo {pages} {1901606} (\bibinfo {year} {2021})}\BibitemShut
  {NoStop}%
\bibitem [{\citenamefont {Bergman}\ \emph {et~al.}(2008)\citenamefont
  {Bergman}, \citenamefont {Wu},\ and\ \citenamefont {Balents}}]{Bergman2008}%
  \BibitemOpen
  \bibfield  {author} {\bibinfo {author} {\bibfnamefont {D.~L.}\ \bibnamefont
  {Bergman}}, \bibinfo {author} {\bibfnamefont {C.}~\bibnamefont {Wu}},\ and\
  \bibinfo {author} {\bibfnamefont {L.}~\bibnamefont {Balents}},\ }\bibfield
  {title} {\bibinfo {title} {Band touching from real-space topology in
  frustrated hopping models},\ }\href
  {https://doi.org/10.1103/PhysRevB.78.125104} {\bibfield  {journal} {\bibinfo
  {journal} {Physical Review B}\ }\textbf {\bibinfo {volume} {78}},\ \bibinfo
  {pages} {125104} (\bibinfo {year} {2008})}\BibitemShut {NoStop}%
\bibitem [{\citenamefont {Hasan}\ and\ \citenamefont {Kane}(2010)}]{Hasan2010}%
  \BibitemOpen
  \bibfield  {author} {\bibinfo {author} {\bibfnamefont {M.~Z.}\ \bibnamefont
  {Hasan}}\ and\ \bibinfo {author} {\bibfnamefont {C.~L.}\ \bibnamefont
  {Kane}},\ }\bibfield  {title} {\bibinfo {title} {Colloquium: {{Topological}}
  insulators},\ }\href {https://doi.org/10.1103/RevModPhys.82.3045} {\bibfield
  {journal} {\bibinfo  {journal} {Reviews of Modern Physics}\ }\textbf
  {\bibinfo {volume} {82}},\ \bibinfo {pages} {3045} (\bibinfo {year}
  {2010})}\BibitemShut {NoStop}%
\bibitem [{\citenamefont {Lu}\ \emph {et~al.}(2014)\citenamefont {Lu},
  \citenamefont {Joannopoulos},\ and\ \citenamefont {Solja{\v
  c}i{\'c}}}]{Lu2014}%
  \BibitemOpen
  \bibfield  {author} {\bibinfo {author} {\bibfnamefont {L.}~\bibnamefont
  {Lu}}, \bibinfo {author} {\bibfnamefont {J.~D.}\ \bibnamefont
  {Joannopoulos}},\ and\ \bibinfo {author} {\bibfnamefont {M.}~\bibnamefont
  {Solja{\v c}i{\'c}}},\ }\bibfield  {title} {\bibinfo {title} {Topological
  photonics},\ }\href {https://doi.org/10.1038/nphoton.2014.248} {\bibfield
  {journal} {\bibinfo  {journal} {Nature Photonics}\ }\textbf {\bibinfo
  {volume} {8}},\ \bibinfo {pages} {821} (\bibinfo {year} {2014})}\BibitemShut
  {NoStop}%
\bibitem [{\citenamefont {Huber}(2016)}]{Huber2016}%
  \BibitemOpen
  \bibfield  {author} {\bibinfo {author} {\bibfnamefont {S.~D.}\ \bibnamefont
  {Huber}},\ }\bibfield  {title} {\bibinfo {title} {Topological mechanics},\
  }\href {https://doi.org/10.1038/nphys3801} {\bibfield  {journal} {\bibinfo
  {journal} {Nature Physics}\ }\textbf {\bibinfo {volume} {12}},\ \bibinfo
  {pages} {621} (\bibinfo {year} {2016})}\BibitemShut {NoStop}%
\bibitem [{\citenamefont {Rhim}\ \emph {et~al.}(2020)\citenamefont {Rhim},
  \citenamefont {Kim},\ and\ \citenamefont {Yang}}]{Rhim_2020}%
  \BibitemOpen
  \bibfield  {author} {\bibinfo {author} {\bibfnamefont {J.-W.}\ \bibnamefont
  {Rhim}}, \bibinfo {author} {\bibfnamefont {K.}~\bibnamefont {Kim}},\ and\
  \bibinfo {author} {\bibfnamefont {B.-J.}\ \bibnamefont {Yang}},\ }\bibfield
  {title} {\bibinfo {title} {Quantum distance and anomalous landau levels of
  flat bands},\ }\href {https://doi.org/10.1038/s41586-020-2540-1} {\bibfield
  {journal} {\bibinfo  {journal} {Nature}\ }\textbf {\bibinfo {volume} {584}},\
  \bibinfo {pages} {59} (\bibinfo {year} {2020})}\BibitemShut {NoStop}%
\bibitem [{\citenamefont {Xia}\ \emph {et~al.}(2018)\citenamefont {Xia},
  \citenamefont {Ramachandran}, \citenamefont {Xia}, \citenamefont {Li},
  \citenamefont {Liu}, \citenamefont {Tang}, \citenamefont {Hu}, \citenamefont
  {Song}, \citenamefont {Xu}, \citenamefont {Leykam}, \citenamefont {Flach},\
  and\ \citenamefont {Chen}}]{Xia2018}%
  \BibitemOpen
  \bibfield  {author} {\bibinfo {author} {\bibfnamefont {S.}~\bibnamefont
  {Xia}}, \bibinfo {author} {\bibfnamefont {A.}~\bibnamefont {Ramachandran}},
  \bibinfo {author} {\bibfnamefont {S.}~\bibnamefont {Xia}}, \bibinfo {author}
  {\bibfnamefont {D.}~\bibnamefont {Li}}, \bibinfo {author} {\bibfnamefont
  {X.}~\bibnamefont {Liu}}, \bibinfo {author} {\bibfnamefont {L.}~\bibnamefont
  {Tang}}, \bibinfo {author} {\bibfnamefont {Y.}~\bibnamefont {Hu}}, \bibinfo
  {author} {\bibfnamefont {D.}~\bibnamefont {Song}}, \bibinfo {author}
  {\bibfnamefont {J.}~\bibnamefont {Xu}}, \bibinfo {author} {\bibfnamefont
  {D.}~\bibnamefont {Leykam}}, \bibinfo {author} {\bibfnamefont
  {S.}~\bibnamefont {Flach}},\ and\ \bibinfo {author} {\bibfnamefont
  {Z.}~\bibnamefont {Chen}},\ }\bibfield  {title} {\bibinfo {title}
  {Unconventional {{Flatband Line States}} in {{Photonic Lieb Lattices}}},\
  }\href {https://doi.org/10.1103/PhysRevLett.121.263902} {\bibfield  {journal}
  {\bibinfo  {journal} {Physical Review Letters}\ }\textbf {\bibinfo {volume}
  {121}},\ \bibinfo {pages} {263902} (\bibinfo {year} {2018})}\BibitemShut
  {NoStop}%
\bibitem [{\citenamefont {Ma}\ \emph {et~al.}(2020)\citenamefont {Ma},
  \citenamefont {Rhim}, \citenamefont {Tang}, \citenamefont {Xia},
  \citenamefont {Wang}, \citenamefont {Zheng}, \citenamefont {Xia},
  \citenamefont {Song}, \citenamefont {Hu}, \citenamefont {Li}, \citenamefont
  {Yang}, \citenamefont {Leykam},\ and\ \citenamefont {Chen}}]{Ma_2020}%
  \BibitemOpen
  \bibfield  {author} {\bibinfo {author} {\bibfnamefont {J.}~\bibnamefont
  {Ma}}, \bibinfo {author} {\bibfnamefont {J.-W.}\ \bibnamefont {Rhim}},
  \bibinfo {author} {\bibfnamefont {L.}~\bibnamefont {Tang}}, \bibinfo {author}
  {\bibfnamefont {S.}~\bibnamefont {Xia}}, \bibinfo {author} {\bibfnamefont
  {H.}~\bibnamefont {Wang}}, \bibinfo {author} {\bibfnamefont {X.}~\bibnamefont
  {Zheng}}, \bibinfo {author} {\bibfnamefont {S.}~\bibnamefont {Xia}}, \bibinfo
  {author} {\bibfnamefont {D.}~\bibnamefont {Song}}, \bibinfo {author}
  {\bibfnamefont {Y.}~\bibnamefont {Hu}}, \bibinfo {author} {\bibfnamefont
  {Y.}~\bibnamefont {Li}}, \bibinfo {author} {\bibfnamefont {B.-J.}\
  \bibnamefont {Yang}}, \bibinfo {author} {\bibfnamefont {D.}~\bibnamefont
  {Leykam}},\ and\ \bibinfo {author} {\bibfnamefont {Z.}~\bibnamefont {Chen}},\
  }\bibfield  {title} {\bibinfo {title} {Direct observation of flatband loop
  states arising from nontrivial real-space topology},\ }\href
  {https://doi.org/10.1103/PhysRevLett.124.183901} {\bibfield  {journal}
  {\bibinfo  {journal} {Phys. Rev. Lett.}\ }\textbf {\bibinfo {volume} {124}},\
  \bibinfo {pages} {183901} (\bibinfo {year} {2020})}\BibitemShut {NoStop}%
\bibitem [{\citenamefont {Xie}\ \emph {et~al.}(2021)\citenamefont {Xie},
  \citenamefont {Song}, \citenamefont {Yan}, \citenamefont {Xia}, \citenamefont
  {Tang}, \citenamefont {Song}, \citenamefont {Rhim},\ and\ \citenamefont
  {Chen}}]{Xie2021}%
  \BibitemOpen
  \bibfield  {author} {\bibinfo {author} {\bibfnamefont {Y.}~\bibnamefont
  {Xie}}, \bibinfo {author} {\bibfnamefont {L.}~\bibnamefont {Song}}, \bibinfo
  {author} {\bibfnamefont {W.}~\bibnamefont {Yan}}, \bibinfo {author}
  {\bibfnamefont {S.}~\bibnamefont {Xia}}, \bibinfo {author} {\bibfnamefont
  {L.}~\bibnamefont {Tang}}, \bibinfo {author} {\bibfnamefont {D.}~\bibnamefont
  {Song}}, \bibinfo {author} {\bibfnamefont {J.-W.}\ \bibnamefont {Rhim}},\
  and\ \bibinfo {author} {\bibfnamefont {Z.}~\bibnamefont {Chen}},\ }\bibfield
  {title} {\bibinfo {title} {Fractal-like photonic lattices and localized
  states arising from singular and nonsingular flatbands},\ }\href@noop {}
  {\bibfield  {journal} {\bibinfo  {journal} {APL Photonics}\ }\textbf
  {\bibinfo {volume} {6}},\ \bibinfo {pages} {116104} (\bibinfo {year}
  {2021})}\BibitemShut {NoStop}%
\bibitem [{\citenamefont {Hanafi}\ \emph {et~al.}(2022)\citenamefont {Hanafi},
  \citenamefont {Menz},\ and\ \citenamefont {Denz}}]{Hanafi-2022}%
  \BibitemOpen
  \bibfield  {author} {\bibinfo {author} {\bibfnamefont {H.}~\bibnamefont
  {Hanafi}}, \bibinfo {author} {\bibfnamefont {P.}~\bibnamefont {Menz}},\ and\
  \bibinfo {author} {\bibfnamefont {C.}~\bibnamefont {Denz}},\ }\bibfield
  {title} {\bibinfo {title} {Localized states emerging from singular and
  nonsingular flat bands in a frustrated fractal-like photonic lattice},\
  }\href {https://doi.org/https://doi.org/10.1002/adom.202102523} {\bibfield
  {journal} {\bibinfo  {journal} {Advanced Optical Materials}\ }\textbf
  {\bibinfo {volume} {10}},\ \bibinfo {pages} {2102523} (\bibinfo {year}
  {2022})}\BibitemShut {NoStop}%
\bibitem [{\citenamefont {Wang}\ \emph {et~al.}(2020)\citenamefont {Wang},
  \citenamefont {Wang}, \citenamefont {Wu}, \citenamefont {Chen},\ and\
  \citenamefont {Wang}}]{Wang2020}%
  \BibitemOpen
  \bibfield  {author} {\bibinfo {author} {\bibfnamefont {Y.-F.}\ \bibnamefont
  {Wang}}, \bibinfo {author} {\bibfnamefont {Y.-Z.}\ \bibnamefont {Wang}},
  \bibinfo {author} {\bibfnamefont {B.}~\bibnamefont {Wu}}, \bibinfo {author}
  {\bibfnamefont {W.}~\bibnamefont {Chen}},\ and\ \bibinfo {author}
  {\bibfnamefont {Y.-S.}\ \bibnamefont {Wang}},\ }\bibfield  {title} {\bibinfo
  {title} {Tunable and {{Active Phononic Crystals}} and {{Metamaterials}}},\
  }\bibfield  {journal} {\bibinfo  {journal} {Applied Mechanics Reviews}\
  }\textbf {\bibinfo {volume} {72}},\ \href {https://doi.org/10.1115/1.4046222}
  {10.1115/1.4046222} (\bibinfo {year} {2020})\BibitemShut {NoStop}%
\bibitem [{\citenamefont {Karki}\ and\ \citenamefont
  {Paulose}(2021)}]{Karki_2021}%
  \BibitemOpen
  \bibfield  {author} {\bibinfo {author} {\bibfnamefont {P.}~\bibnamefont
  {Karki}}\ and\ \bibinfo {author} {\bibfnamefont {J.}~\bibnamefont
  {Paulose}},\ }\bibfield  {title} {\bibinfo {title} {Stopping and reversing
  sound via dynamic dispersion tuning in a phononic metamaterial},\ }\href
  {https://doi.org/10.1103/PhysRevApplied.15.034083} {\bibfield  {journal}
  {\bibinfo  {journal} {Phys. Rev. Applied}\ }\textbf {\bibinfo {volume}
  {15}},\ \bibinfo {pages} {034083} (\bibinfo {year} {2021})}\BibitemShut
  {NoStop}%
\bibitem [{\citenamefont {Cha}\ and\ \citenamefont {Daraio}(2018)}]{Cha_2018}%
  \BibitemOpen
  \bibfield  {author} {\bibinfo {author} {\bibfnamefont {J.}~\bibnamefont
  {Cha}}\ and\ \bibinfo {author} {\bibfnamefont {C.}~\bibnamefont {Daraio}},\
  }\bibfield  {title} {{\selectlanguage {English}\bibinfo {title} {Electrical
  tuning of elastic wave propagation in nanomechanical lattices at {{MHz}}
  frequencies}},\ }\href {https://doi.org/10.1038/s41565-018-0252-6} {\bibfield
   {journal} {\bibinfo  {journal} {Nature Nanotechnology}\ }\textbf {\bibinfo
  {volume} {13}},\ \bibinfo {pages} {1016} (\bibinfo {year}
  {2018})}\BibitemShut {NoStop}%
\bibitem [{\citenamefont {Mei}\ \emph {et~al.}(2018)\citenamefont {Mei},
  \citenamefont {Lee}, \citenamefont {Xu},\ and\ \citenamefont
  {Feng}}]{Mei_2018}%
  \BibitemOpen
  \bibfield  {author} {\bibinfo {author} {\bibfnamefont {T.}~\bibnamefont
  {Mei}}, \bibinfo {author} {\bibfnamefont {J.}~\bibnamefont {Lee}}, \bibinfo
  {author} {\bibfnamefont {Y.}~\bibnamefont {Xu}},\ and\ \bibinfo {author}
  {\bibfnamefont {P.~X.-L.}\ \bibnamefont {Feng}},\ }\bibfield  {title}
  {\bibinfo {title} {Frequency tuning of graphene nanoelectromechanical
  resonators via electrostatic gating},\ }\href@noop {} {\bibfield  {journal}
  {\bibinfo  {journal} {Micromachines}\ }\textbf {\bibinfo {volume} {9}},\
  \bibinfo {pages} {312} (\bibinfo {year} {2018})}\BibitemShut {NoStop}%
\bibitem [{\citenamefont {Blaikie}\ \emph {et~al.}(2019)\citenamefont
  {Blaikie}, \citenamefont {Miller},\ and\ \citenamefont
  {Alem{\'a}n}}]{Blaikie_2019}%
  \BibitemOpen
  \bibfield  {author} {\bibinfo {author} {\bibfnamefont {A.}~\bibnamefont
  {Blaikie}}, \bibinfo {author} {\bibfnamefont {D.}~\bibnamefont {Miller}},\
  and\ \bibinfo {author} {\bibfnamefont {B.~J.}\ \bibnamefont {Alem{\'a}n}},\
  }\bibfield  {title} {{\selectlanguage {English}\bibinfo {title} {A fast and
  sensitive room-temperature graphene nanomechanical bolometer}},\ }\href
  {https://doi.org/10.1038/s41467-019-12562-2} {\bibfield  {journal} {\bibinfo
  {journal} {Nature Communications}\ }\textbf {\bibinfo {volume} {10}},\
  \bibinfo {pages} {1} (\bibinfo {year} {2019})}\BibitemShut {NoStop}%
\bibitem [{\citenamefont {Zande}\ \emph {et~al.}(2010)\citenamefont {Zande},
  \citenamefont {Barton}, \citenamefont {Alden}, \citenamefont {Ruiz-Vargas},
  \citenamefont {Whitney}, \citenamefont {Pham}, \citenamefont {Park},
  \citenamefont {Parpia}, \citenamefont {Craighead},\ and\ \citenamefont
  {McEuen}}]{Zande_2010}%
  \BibitemOpen
  \bibfield  {author} {\bibinfo {author} {\bibfnamefont {A.~M. v.~d.}\
  \bibnamefont {Zande}}, \bibinfo {author} {\bibfnamefont {R.~A.}\ \bibnamefont
  {Barton}}, \bibinfo {author} {\bibfnamefont {J.~S.}\ \bibnamefont {Alden}},
  \bibinfo {author} {\bibfnamefont {C.~S.}\ \bibnamefont {Ruiz-Vargas}},
  \bibinfo {author} {\bibfnamefont {W.~S.}\ \bibnamefont {Whitney}}, \bibinfo
  {author} {\bibfnamefont {P.~H.~Q.}\ \bibnamefont {Pham}}, \bibinfo {author}
  {\bibfnamefont {J.}~\bibnamefont {Park}}, \bibinfo {author} {\bibfnamefont
  {J.~M.}\ \bibnamefont {Parpia}}, \bibinfo {author} {\bibfnamefont {H.~G.}\
  \bibnamefont {Craighead}},\ and\ \bibinfo {author} {\bibfnamefont {P.~L.}\
  \bibnamefont {McEuen}},\ }\bibfield  {title} {\bibinfo {title} {Large-scale
  arrays of single-layer graphene resonators},\ }\bibfield  {booktitle} {\emph
  {\bibinfo {booktitle} {Nano Letters}},\ }\href
  {https://doi.org/10.1021/nl102713c} {\bibfield  {journal} {\bibinfo
  {journal} {Nano Letters}\ }\textbf {\bibinfo {volume} {10}},\ \bibinfo
  {pages} {4869} (\bibinfo {year} {2010})}\BibitemShut {NoStop}%
\bibitem [{\citenamefont {Timoshenko}\ and\ \citenamefont
  {{Woinowsky-Krieger}}(1959)}]{Timoshenkotheory_1959}%
  \BibitemOpen
  \bibfield  {author} {\bibinfo {author} {\bibfnamefont {S.}~\bibnamefont
  {Timoshenko}}\ and\ \bibinfo {author} {\bibfnamefont {S.}~\bibnamefont
  {{Woinowsky-Krieger}}},\ }\href@noop {} {\emph {\bibinfo {title} {Theory of
  Plates and Shells}}},\ Engineering Mechanics Series\ (\bibinfo  {publisher}
  {{McGraw-Hill}},\ \bibinfo {year} {1959})\BibitemShut {NoStop}%
\bibitem [{\citenamefont {Sweers}(2001)}]{Sweers_2001}%
  \BibitemOpen
  \bibfield  {author} {\bibinfo {author} {\bibfnamefont {G.}~\bibnamefont
  {Sweers}},\ }\bibfield  {title} {\bibinfo {title} {When is the first
  eigenfunction for the clamped plate equation of fixed sign?},\ }\href
  {http://eudml.org/doc/121016} {\bibfield  {journal} {\bibinfo  {journal}
  {Electronic Journal of Differential Equations (EJDE) [electronic only]}\ ,\
  \bibinfo {pages} {285}} (\bibinfo {year} {2001})}\BibitemShut {NoStop}%
\bibitem [{\citenamefont {Brown}\ \emph {et~al.}(1999)\citenamefont {Brown},
  \citenamefont {Davies}, \citenamefont {Jimack},\ and\ \citenamefont
  {Mihajlovi'c}}]{Brown_1999}%
  \BibitemOpen
  \bibfield  {author} {\bibinfo {author} {\bibfnamefont {B.~M.}\ \bibnamefont
  {Brown}}, \bibinfo {author} {\bibfnamefont {E.~B.}\ \bibnamefont {Davies}},
  \bibinfo {author} {\bibfnamefont {P.~K.}\ \bibnamefont {Jimack}},\ and\
  \bibinfo {author} {\bibfnamefont {M.~D.}\ \bibnamefont {Mihajlovi'c}},\
  }\href@noop {} {\bibinfo {title} {On the accurate finite element solution of
  a class of fourth order eigenvalue problems}} (\bibinfo {year} {1999}),\
  \Eprint {https://arxiv.org/abs/math/9905038} {arXiv:math/9905038 [math.SP]}
  \BibitemShut {NoStop}%
\bibitem [{\citenamefont {Keil}\ \emph {et~al.}(2016)\citenamefont {Keil},
  \citenamefont {Poli}, \citenamefont {Heinrich}, \citenamefont {Arkinstall},
  \citenamefont {Weihs}, \citenamefont {Schomerus},\ and\ \citenamefont
  {Szameit}}]{Keil2016}%
  \BibitemOpen
  \bibfield  {author} {\bibinfo {author} {\bibfnamefont {R.}~\bibnamefont
  {Keil}}, \bibinfo {author} {\bibfnamefont {C.}~\bibnamefont {Poli}}, \bibinfo
  {author} {\bibfnamefont {M.}~\bibnamefont {Heinrich}}, \bibinfo {author}
  {\bibfnamefont {J.}~\bibnamefont {Arkinstall}}, \bibinfo {author}
  {\bibfnamefont {G.}~\bibnamefont {Weihs}}, \bibinfo {author} {\bibfnamefont
  {H.}~\bibnamefont {Schomerus}},\ and\ \bibinfo {author} {\bibfnamefont
  {A.}~\bibnamefont {Szameit}},\ }\bibfield  {title} {\bibinfo {title}
  {Universal {{Sign Control}} of {{Coupling}} in {{Tight-Binding Lattices}}},\
  }\href {https://doi.org/10.1103/PhysRevLett.116.213901} {\bibfield  {journal}
  {\bibinfo  {journal} {Physical Review Letters}\ }\textbf {\bibinfo {volume}
  {116}},\ \bibinfo {pages} {213901} (\bibinfo {year} {2016})}\BibitemShut
  {NoStop}%
\bibitem [{\citenamefont {Castro~Neto}\ \emph {et~al.}(2009)\citenamefont
  {Castro~Neto}, \citenamefont {Guinea}, \citenamefont {Peres}, \citenamefont
  {Novoselov},\ and\ \citenamefont {Geim}}]{CastroNeto2009}%
  \BibitemOpen
  \bibfield  {author} {\bibinfo {author} {\bibfnamefont {a.~H.}\ \bibnamefont
  {Castro~Neto}}, \bibinfo {author} {\bibfnamefont {F.}~\bibnamefont {Guinea}},
  \bibinfo {author} {\bibfnamefont {N.~M.~R.}\ \bibnamefont {Peres}}, \bibinfo
  {author} {\bibfnamefont {K.~S.}\ \bibnamefont {Novoselov}},\ and\ \bibinfo
  {author} {\bibfnamefont {a.~K.}\ \bibnamefont {Geim}},\ }\bibfield  {title}
  {\bibinfo {title} {The electronic properties of graphene},\ }\href
  {https://doi.org/10.1103/RevModPhys.81.109} {\bibfield  {journal} {\bibinfo
  {journal} {Reviews of Modern Physics}\ }\textbf {\bibinfo {volume} {81}},\
  \bibinfo {pages} {109} (\bibinfo {year} {2009})},\ \Eprint
  {https://arxiv.org/abs/0709.1163} {arXiv:0709.1163} \BibitemShut {NoStop}%
\bibitem [{\citenamefont {Berry}(1989)}]{Berry1989}%
  \BibitemOpen
  \bibfield  {author} {\bibinfo {author} {\bibfnamefont {M.~V.}\ \bibnamefont
  {Berry}},\ }\bibfield  {title} {\bibinfo {title} {The quantum phase, five
  years after},\ }in\ \href@noop {} {\emph {\bibinfo {booktitle} {Geometric
  phases in physics}}},\ Vol.~\bibinfo {volume} {5},\ \bibinfo {editor} {edited
  by\ \bibinfo {editor} {\bibfnamefont {A.}~\bibnamefont {Shapere}}\ and\
  \bibinfo {editor} {\bibfnamefont {F.}~\bibnamefont {Wilczek}}}\ (\bibinfo
  {publisher} {World Scientific},\ \bibinfo {address} {Singapore},\ \bibinfo
  {year} {1989})\BibitemShut {NoStop}%
\bibitem [{\citenamefont {Xiao}\ \emph {et~al.}(2010)\citenamefont {Xiao},
  \citenamefont {Chang},\ and\ \citenamefont {Niu}}]{Xiao2010a}%
  \BibitemOpen
  \bibfield  {author} {\bibinfo {author} {\bibfnamefont {D.}~\bibnamefont
  {Xiao}}, \bibinfo {author} {\bibfnamefont {M.-C.}\ \bibnamefont {Chang}},\
  and\ \bibinfo {author} {\bibfnamefont {Q.}~\bibnamefont {Niu}},\ }\bibfield
  {title} {\bibinfo {title} {Berry phase effects on electronic properties},\
  }\href {https://doi.org/10.1103/RevModPhys.82.1959} {\bibfield  {journal}
  {\bibinfo  {journal} {Reviews of Modern Physics}\ }\textbf {\bibinfo {volume}
  {82}},\ \bibinfo {pages} {1959} (\bibinfo {year} {2010})}\BibitemShut
  {NoStop}%
\bibitem [{\citenamefont {Tang}\ \emph {et~al.}(2020)\citenamefont {Tang},
  \citenamefont {Song}, \citenamefont {Xia}, \citenamefont {Xia}, \citenamefont
  {Ma}, \citenamefont {Yan}, \citenamefont {Hu}, \citenamefont {Xu},
  \citenamefont {Leykam},\ and\ \citenamefont {Chen}}]{Tang2020}%
  \BibitemOpen
  \bibfield  {author} {\bibinfo {author} {\bibfnamefont {L.}~\bibnamefont
  {Tang}}, \bibinfo {author} {\bibfnamefont {D.}~\bibnamefont {Song}}, \bibinfo
  {author} {\bibfnamefont {S.}~\bibnamefont {Xia}}, \bibinfo {author}
  {\bibfnamefont {S.}~\bibnamefont {Xia}}, \bibinfo {author} {\bibfnamefont
  {J.}~\bibnamefont {Ma}}, \bibinfo {author} {\bibfnamefont {W.}~\bibnamefont
  {Yan}}, \bibinfo {author} {\bibfnamefont {Y.}~\bibnamefont {Hu}}, \bibinfo
  {author} {\bibfnamefont {J.}~\bibnamefont {Xu}}, \bibinfo {author}
  {\bibfnamefont {D.}~\bibnamefont {Leykam}},\ and\ \bibinfo {author}
  {\bibfnamefont {Z.}~\bibnamefont {Chen}},\ }\bibfield  {title} {\bibinfo
  {title} {Photonic flat-band lattices and unconventional light localization},\
  }\bibfield  {journal} {\bibinfo  {journal} {Nanophotonics}\ }\textbf
  {\bibinfo {volume} {-1}},\ \href {https://doi.org/10.1515/nanoph-2020-0043}
  {10.1515/nanoph-2020-0043} (\bibinfo {year} {2020})\BibitemShut {NoStop}%
\bibitem [{\citenamefont {Ankit}\ \emph {et~al.}(2021)\citenamefont {Ankit},
  \citenamefont {Krisnadi}, \citenamefont {Pethe}, \citenamefont {Lim},
  \citenamefont {Kulkarni}, \citenamefont {Accoto},\ and\ \citenamefont
  {Mathews}}]{Ankit_2021}%
  \BibitemOpen
  \bibfield  {author} {\bibinfo {author} {\bibnamefont {Ankit}}, \bibinfo
  {author} {\bibfnamefont {F.}~\bibnamefont {Krisnadi}}, \bibinfo {author}
  {\bibfnamefont {S.}~\bibnamefont {Pethe}}, \bibinfo {author} {\bibfnamefont
  {K.~J.~R.}\ \bibnamefont {Lim}}, \bibinfo {author} {\bibfnamefont {M.~R.}\
  \bibnamefont {Kulkarni}}, \bibinfo {author} {\bibfnamefont {D.}~\bibnamefont
  {Accoto}},\ and\ \bibinfo {author} {\bibfnamefont {N.}~\bibnamefont
  {Mathews}},\ }\bibfield  {title} {\bibinfo {title} {Mxene incorporated
  polymeric hybrids for stiffness modulation in printed adaptive surfaces},\
  }\href {https://doi.org/https://doi.org/10.1016/j.nanoen.2021.106548}
  {\bibfield  {journal} {\bibinfo  {journal} {Nano Energy}\ }\textbf {\bibinfo
  {volume} {90}},\ \bibinfo {pages} {106548} (\bibinfo {year}
  {2021})}\BibitemShut {NoStop}%
\bibitem [{\citenamefont {Siriwardane}\ \emph {et~al.}(2018)\citenamefont
  {Siriwardane}, \citenamefont {Karki}, \citenamefont {Sevik},\ and\
  \citenamefont {Çakır}}]{Siriwardane_2018}%
  \BibitemOpen
  \bibfield  {author} {\bibinfo {author} {\bibfnamefont {E.~M.}\ \bibnamefont
  {Siriwardane}}, \bibinfo {author} {\bibfnamefont {P.}~\bibnamefont {Karki}},
  \bibinfo {author} {\bibfnamefont {C.}~\bibnamefont {Sevik}},\ and\ \bibinfo
  {author} {\bibfnamefont {D.}~\bibnamefont {Çakır}},\ }\bibfield  {title}
  {\bibinfo {title} {Electronic and mechanical properties of stiff rhenium
  carbide monolayers: A first-principles investigation},\ }\href
  {https://doi.org/https://doi.org/10.1016/j.apsusc.2018.07.058} {\bibfield
  {journal} {\bibinfo  {journal} {Applied Surface Science}\ }\textbf {\bibinfo
  {volume} {458}},\ \bibinfo {pages} {762} (\bibinfo {year}
  {2018})}\BibitemShut {NoStop}%
\bibitem [{\citenamefont {Mera}\ and\ \citenamefont
  {Mitscherling}(2022)}]{Mera2022}%
  \BibitemOpen
  \bibfield  {author} {\bibinfo {author} {\bibfnamefont {B.}~\bibnamefont
  {Mera}}\ and\ \bibinfo {author} {\bibfnamefont {J.}~\bibnamefont
  {Mitscherling}},\ }\href {https://doi.org/10.48550/arXiv.2205.07900}
  {\bibinfo {title} {Nontrivial quantum geometry of degenerate flat bands}}
  (\bibinfo {year} {2022}),\ \Eprint {https://arxiv.org/abs/2205.07900}
  {arXiv:2205.07900 [cond-mat, physics:math-ph]} \BibitemShut {NoStop}%
\bibitem [{\citenamefont {Yang}\ \emph {et~al.}(2017)\citenamefont {Yang},
  \citenamefont {Gao}, \citenamefont {Yang},\ and\ \citenamefont
  {Zhang}}]{Yang2017}%
  \BibitemOpen
  \bibfield  {author} {\bibinfo {author} {\bibfnamefont {Z.}~\bibnamefont
  {Yang}}, \bibinfo {author} {\bibfnamefont {F.}~\bibnamefont {Gao}}, \bibinfo
  {author} {\bibfnamefont {Y.}~\bibnamefont {Yang}},\ and\ \bibinfo {author}
  {\bibfnamefont {B.}~\bibnamefont {Zhang}},\ }\bibfield  {title} {\bibinfo
  {title} {Strain-induced gauge field and landau levels in acoustic
  structures},\ }\href@noop {} {\bibfield  {journal} {\bibinfo  {journal}
  {Physical Review Letters}\ }\textbf {\bibinfo {volume} {118}},\ \bibinfo
  {pages} {194301} (\bibinfo {year} {2017})}\BibitemShut {NoStop}%
\bibitem [{\citenamefont {Abbaszadeh}\ \emph {et~al.}(2017)\citenamefont
  {Abbaszadeh}, \citenamefont {Souslov}, \citenamefont {Paulose}, \citenamefont
  {Schomerus},\ and\ \citenamefont {Vitelli}}]{Abbaszadeh2016}%
  \BibitemOpen
  \bibfield  {author} {\bibinfo {author} {\bibfnamefont {H.}~\bibnamefont
  {Abbaszadeh}}, \bibinfo {author} {\bibfnamefont {A.}~\bibnamefont {Souslov}},
  \bibinfo {author} {\bibfnamefont {J.}~\bibnamefont {Paulose}}, \bibinfo
  {author} {\bibfnamefont {H.}~\bibnamefont {Schomerus}},\ and\ \bibinfo
  {author} {\bibfnamefont {V.}~\bibnamefont {Vitelli}},\ }\bibfield  {title}
  {\bibinfo {title} {Sonic {{Landau Levels}} and {{Synthetic Gauge Fields}} in
  {{Mechanical Metamaterials}}},\ }\href
  {https://doi.org/10.1103/PhysRevLett.119.195502} {\bibfield  {journal}
  {\bibinfo  {journal} {Physical Review Letters}\ }\textbf {\bibinfo {volume}
  {119}},\ \bibinfo {pages} {195502} (\bibinfo {year} {2017})},\ \Eprint
  {https://arxiv.org/abs/1610.06406} {arXiv:1610.06406} \BibitemShut {NoStop}%
\bibitem [{\citenamefont {Brendel}\ \emph {et~al.}(2017)\citenamefont
  {Brendel}, \citenamefont {Peano}, \citenamefont {Painter},\ and\
  \citenamefont {Marquardt}}]{Brendel2017}%
  \BibitemOpen
  \bibfield  {author} {\bibinfo {author} {\bibfnamefont {C.}~\bibnamefont
  {Brendel}}, \bibinfo {author} {\bibfnamefont {V.}~\bibnamefont {Peano}},
  \bibinfo {author} {\bibfnamefont {O.~J.}\ \bibnamefont {Painter}},\ and\
  \bibinfo {author} {\bibfnamefont {F.}~\bibnamefont {Marquardt}},\ }\bibfield
  {title} {\bibinfo {title} {Pseudomagnetic fields for sound at the
  nanoscale},\ }\href {https://doi.org/10.1073/pnas.1615503114} {\bibfield
  {journal} {\bibinfo  {journal} {Proceedings of the National Academy of
  Sciences}\ }\textbf {\bibinfo {volume} {114}},\ \bibinfo {pages} {E3390}
  (\bibinfo {year} {2017})}\BibitemShut {NoStop}%
\bibitem [{\citenamefont {Wen}\ \emph {et~al.}(2019)\citenamefont {Wen},
  \citenamefont {Qiu}, \citenamefont {Qi}, \citenamefont {Ye}, \citenamefont
  {Ke}, \citenamefont {Zhang},\ and\ \citenamefont {Liu}}]{Wen2019}%
  \BibitemOpen
  \bibfield  {author} {\bibinfo {author} {\bibfnamefont {X.}~\bibnamefont
  {Wen}}, \bibinfo {author} {\bibfnamefont {C.}~\bibnamefont {Qiu}}, \bibinfo
  {author} {\bibfnamefont {Y.}~\bibnamefont {Qi}}, \bibinfo {author}
  {\bibfnamefont {L.}~\bibnamefont {Ye}}, \bibinfo {author} {\bibfnamefont
  {M.}~\bibnamefont {Ke}}, \bibinfo {author} {\bibfnamefont {F.}~\bibnamefont
  {Zhang}},\ and\ \bibinfo {author} {\bibfnamefont {Z.}~\bibnamefont {Liu}},\
  }\bibfield  {title} {\bibinfo {title} {Acoustic landau quantization and
  quantum-hall-like edge states},\ }\href@noop {} {\bibfield  {journal}
  {\bibinfo  {journal} {Nature Physics}\ }\textbf {\bibinfo {volume} {15}},\
  \bibinfo {pages} {352} (\bibinfo {year} {2019})}\BibitemShut {NoStop}%
\bibitem [{\citenamefont {Yan}\ \emph {et~al.}(2021)\citenamefont {Yan},
  \citenamefont {Deng}, \citenamefont {Huang}, \citenamefont {Wu},
  \citenamefont {Yang}, \citenamefont {Lu}, \citenamefont {Li},\ and\
  \citenamefont {Liu}}]{Yan2021}%
  \BibitemOpen
  \bibfield  {author} {\bibinfo {author} {\bibfnamefont {M.}~\bibnamefont
  {Yan}}, \bibinfo {author} {\bibfnamefont {W.}~\bibnamefont {Deng}}, \bibinfo
  {author} {\bibfnamefont {X.}~\bibnamefont {Huang}}, \bibinfo {author}
  {\bibfnamefont {Y.}~\bibnamefont {Wu}}, \bibinfo {author} {\bibfnamefont
  {Y.}~\bibnamefont {Yang}}, \bibinfo {author} {\bibfnamefont {J.}~\bibnamefont
  {Lu}}, \bibinfo {author} {\bibfnamefont {F.}~\bibnamefont {Li}},\ and\
  \bibinfo {author} {\bibfnamefont {Z.}~\bibnamefont {Liu}},\ }\bibfield
  {title} {\bibinfo {title} {Pseudomagnetic fields enabled manipulation of
  on-chip elastic waves},\ }\href@noop {} {\bibfield  {journal} {\bibinfo
  {journal} {Physical Review Letters}\ }\textbf {\bibinfo {volume} {127}},\
  \bibinfo {pages} {136401} (\bibinfo {year} {2021})}\BibitemShut {NoStop}%
\bibitem [{\citenamefont {Rugar}\ and\ \citenamefont
  {Gr{\"u}tter}(1991)}]{Rugar1991}%
  \BibitemOpen
  \bibfield  {author} {\bibinfo {author} {\bibfnamefont {D.}~\bibnamefont
  {Rugar}}\ and\ \bibinfo {author} {\bibfnamefont {P.}~\bibnamefont
  {Gr{\"u}tter}},\ }\bibfield  {title} {\bibinfo {title} {Mechanical parametric
  amplification and thermomechanical noise squeezing},\ }\href
  {https://doi.org/10.1103/PhysRevLett.67.699} {\bibfield  {journal} {\bibinfo
  {journal} {Physical Review Letters}\ }\textbf {\bibinfo {volume} {67}},\
  \bibinfo {pages} {699} (\bibinfo {year} {1991})}\BibitemShut {NoStop}%
\bibitem [{\citenamefont {Fruchart}\ \emph {et~al.}(2020)\citenamefont
  {Fruchart}, \citenamefont {Zhou},\ and\ \citenamefont
  {Vitelli}}]{Fruchart2020}%
  \BibitemOpen
  \bibfield  {author} {\bibinfo {author} {\bibfnamefont {M.}~\bibnamefont
  {Fruchart}}, \bibinfo {author} {\bibfnamefont {Y.}~\bibnamefont {Zhou}},\
  and\ \bibinfo {author} {\bibfnamefont {V.}~\bibnamefont {Vitelli}},\
  }\bibfield  {title} {\bibinfo {title} {Dualities and non-{{Abelian}}
  mechanics},\ }\href {https://doi.org/10.1038/s41586-020-1932-6} {\bibfield
  {journal} {\bibinfo  {journal} {Nature}\ }\textbf {\bibinfo {volume} {577}},\
  \bibinfo {pages} {636} (\bibinfo {year} {2020})}\BibitemShut {NoStop}%
\bibitem [{\citenamefont {Matlack}\ \emph {et~al.}(2018)\citenamefont
  {Matlack}, \citenamefont {{Serra-Garcia}}, \citenamefont {Palermo},
  \citenamefont {Huber},\ and\ \citenamefont {Daraio}}]{Matlack2018a}%
  \BibitemOpen
  \bibfield  {author} {\bibinfo {author} {\bibfnamefont {K.~H.}\ \bibnamefont
  {Matlack}}, \bibinfo {author} {\bibfnamefont {M.}~\bibnamefont
  {{Serra-Garcia}}}, \bibinfo {author} {\bibfnamefont {A.}~\bibnamefont
  {Palermo}}, \bibinfo {author} {\bibfnamefont {S.~D.}\ \bibnamefont {Huber}},\
  and\ \bibinfo {author} {\bibfnamefont {C.}~\bibnamefont {Daraio}},\
  }\bibfield  {title} {\bibinfo {title} {Designing perturbative metamaterials
  from discrete models},\ }\href {https://doi.org/10.1038/s41563-017-0003-3}
  {\bibfield  {journal} {\bibinfo  {journal} {Nature Materials}\ }\textbf
  {\bibinfo {volume} {17}},\ \bibinfo {pages} {323} (\bibinfo {year}
  {2018})}\BibitemShut {NoStop}%
\end{thebibliography}
